\RequirePackage[mathlines,running]{lineno}
\documentclass[twocolumn,trackchanges]{aastex62}

\usepackage[normalem]{ulem}

\usepackage{lineno}
\usepackage{amssymb}
\usepackage{array,multirow}
\usepackage{comment}
\usepackage{enumerate}
\usepackage{bm}

\usepackage[nolist,nohyperlinks]{acronym} 

\usepackage{amsmath}
\DeclareMathOperator*{\argmax}{arg\,max}

\newcommand{\Ldata}[1]{\ensuremath{L_{\mathrm{#1}}}} 
\newcommand{\Udata}[1]{\ensuremath{U_{\mathrm{#1}}}} 
\newcommand{\Ddata}{\ensuremath{D}} 

\newcommand{\Simul}[1]{\ensuremath{\mathcal{S}_{\mathrm{#1}}}}

\newcommand{\PhysModel}{\ensuremath{\mathbf{p}}}
\newcommand{\MlModel}{\ensuremath{\Gamma}}

\newcommand{\ClassAcc}{\ensuremath{\mathrm{CA}}}
\newcommand{\perClassAcc}{\ensuremath{\mathrm{PCA}}}
\newcommand{\RegrAcc}{\ensuremath{\mathrm{RA}}}

\newcommand{\secref}[1]{\S\ref{#1}}

\newcommand{\psycris}{\ensuremath{\texttt{psy-cris}}}
\newcommand{\posydon}{\ensuremath{\texttt{POSYDON}}}
\newcommand{\mesa}{\ensuremath{\texttt{MESA}}}

\newcommand{\CIERA}{Center for Interdisciplinary Exploration and Research in Astrophysics (CIERA),1800 Sherman, Evanston, IL 60201, USA}

\newcommand{\RV}{\ifmmode {{\rm RV}}\else RV \fi}

\submitjournal{ApJ}

\shorttitle{Active Learning for Population Synthesis}

\turnoffeditone

\begin{document}

\begin{acronym}[MPC]
\acro{BBH}{binary black hole}
\acro{BH}{black hole}
\acro{NS}{neutron star}
\acro{CO}{compact object}

\acro{ZAMS}{Zero Age Main Sequence}

\acro{CE}{common envelope}
\acro{SNe}{supernova}

\acro{AL}{active learning}
\acro{ML}{machine learning}
\acro{RBF}{radial basis function}
\acro{PTMCMC}{parallel tempered Markov-chain Monte Carlo}

\acro{mesa}[\ensuremath{\texttt{MESA}}]{Modules for Experiments in Stellar Astrophysics}
\acro{posydon}[\ensuremath{\texttt{POSYDON}}]{POpulation SYnthesis  with Detailed binary-evolution simulatiONs}
\acro{bse}[\ensuremath{\texttt{BSE}}]{Binary Stellar Evolution}
\acro{sse}[SSE]{single star evolution}

\acro{3D}{three-dimensional}
\acro{1D}{one-dimensional}
\end{acronym}

\title{Active Learning for Computationally Efficient Distribution of Binary Evolution Simulations}

\author[0000-0003-4474-6528]{Kyle~Akira~Rocha}
\affiliation{ \CIERA{} }
\email{kylerocha2024@u.northwestern.edu}

\author[0000-0001-5261-3923]{Jeff~J.~Andrews}
\affiliation{ \CIERA{} }
\affiliation{Department of Physics, University of Florida, 2001 Museum Rd, Gainesville, FL 32611, USA}

\author[0000-0003-3870-7215]{Christopher~P.~L.~Berry}
\affiliation{ \CIERA{} }
\affiliation{ Institute for Gravitational Research, University of Glasgow, Kelvin Building, University Avenue, Glasgow, G12 8QQ, Scotland}

\author[0000-0002-2077-4914]{Zoheyr Doctor}
\affiliation{ \CIERA{} }

\author[0000-0003-4554-0070]{Aggelos~K.~Katsagelos}
\affiliation{Department of Electrical and Computer Engineering, Northwestern University, Evanston, IL, USA}

\author[0000-0002-5949-8662]{Juan~Gabriel~Serra~Pérez}
\affiliation{Department of Electrical and Computer Engineering, Northwestern University, Evanston, IL, USA}

\author[0000-0002-0338-8181]{Pablo Marchant}
\affiliation{ \CIERA{} }
\affiliation{Institute of Astronomy, KU Leuven, Celestijnenlaan 200D, B-3001, Leuven, Belgium}

\author[0000-0001-9236-5469]{Vicky Kalogera}
\affiliation{ \CIERA{} }

\author[0000-0002-0403-4211]{Scott\,Coughlin}
\affiliation{ \CIERA{} }

\author[0000-0002-3439-0321]{Simone\,S.\,Bavera}
\affiliation{Département d’Astronomie, Université de Genève, Chemin Pegasi 51, CH-1290 Versoix, Switzerland}

\author[0000-0002-4442-5700]{Aaron\,Dotter}
\affiliation{ \CIERA{} }

\author[0000-0003-1474-1523]{Tassos\,Fragos}
\affiliation{Département d’Astronomie, Université de Genève, Chemin Pegasi 51, CH-1290 Versoix, Switzerland}

\author[0000-0003-3684-964X]{Konstantinos\,Kovlakas}
\affiliation{Département d’Astronomie, Université de Genève, Chemin Pegasi 51, CH-1290 Versoix, Switzerland}

\author[0000-0003-4260-960X]{Devina\,Misra}
\affiliation{Département d’Astronomie, Université de Genève, Chemin Pegasi 51, CH-1290 Versoix, Switzerland}

\author[0000-0002-0031-3029]{Zepei\,Xing}
\affiliation{Département d’Astronomie, Université de Genève, Chemin Pegasi 51, CH-1290 Versoix, Switzerland}

\author[0000-0002-7464-498X]{Emmanouil\,Zapartas}
\affiliation{Département d’Astronomie, Université de Genève, Chemin Pegasi 51, CH-1290 Versoix, Switzerland}
\affiliation{IAASARS, National Observatory of Athens, Vas. Pavlou and I. Metaxa, Penteli, 15236, Greece}

\begin{abstract}
Binary stars undergo a variety of interactions and evolutionary phases, critical for predicting and explaining observations. 
Binary population synthesis with full simulation of stellar structure and evolution is computationally expensive, requiring a large number of mass-transfer sequences. 
The recently developed binary population synthesis code \posydon{} incorporates grids of \mesa{} binary-star simulations that are interpolated to model large-scale populations of massive binaries. The traditional method of computing a high-density rectilinear grid of simulations is not scalable for higher-dimension grids, accounting for a range of metallicities, rotation and eccentricity. 
We present a new \emph{active learning} algorithm, \psycris{}, which uses machine learning in the data-gathering process to adaptively and iteratively target simulations to run, resulting in a custom, high-performance training set. 
We test \psycris{} on a toy problem and find the resulting training sets require fewer simulations for accurate classification and regression than either regular or randomly sampled grids. 
We further apply \psycris{} to the target problem of building a dynamic grid of \mesa{} simulations, and we demonstrate that, even without fine tuning, a simulation set of only $\sim 1/4$ the size of a rectilinear grid is sufficient to achieve the same classification accuracy. 
We anticipate further gains when algorithmic parameters are optimized for the targeted application. 
We find that optimizing for classification only may lead to performance losses in regression, and vice versa. 
Lowering the computational cost of producing grids will enable new population synthesis codes such as \posydon{} to cover more input parameters while preserving interpolation accuracies.
\end{abstract}


\section{Introduction}
\label{sec:intro} 

Theoretical studies in astronomy often include simulations of physical processes that are not well constrained or understood with the purpose of making predictions and comparisons against observations.
To account for model or physical uncertainties, simulations are carried out with various combinations of unconstrained parameters to explore the space of possible outcomes.
However, in many cases, the computational cost per simulation is large and limits the exploration of the parameter space.
For example, \ac{3D} hydrodynamic \ac{SNe} simulations can take up to tens of millions of CPU-hours per simulation on current facilities \citep[e.g.,][]{2017MNRAS.472..491M,2019MNRAS.482..351V,2021ApJ...915...28B}.
Large scale cosmological simulations such as those from the Illustris project performed with the \texttt{AREPO} code \citep{2010MNRAS.401..791S}, demand similarly high computational costs \citep{2018MNRAS.475..624N,2021ApJ...906..129W}.
Another area where large computational resources are expended is in modeling single and binary stars with \ac{1D} stellar structure and evolution codes such as \acl{mesa} \citep[\acsu{mesa};][]{2011ApJS..192....3P,2013ApJS..208....4P,2015ApJS..220...15P,2018ApJS..234...34P,2019ApJS..243...10P} especially when considering galactic populations.
Large scale computing is not only a significant investment of time and financial resources, but also comes with a carbon footprint, hence optimization also minimizes our environmental impact \citep{2020NatAs...4..819P}.
In this work, we aim to reduce the cost of producing data sets of binary evolution simulations for large populations through \ac{AL}.

Binary population synthesis (hereafter population synthesis) codes simulate large numbers of binary star systems and their interactions to compare their statistics against observations \citep[e.g.,][]{2020RAA....20..161H}.
While population synthesis has broad applicability, there are large computational hurdles to overcome when modeling binary populations. 
Not only does one have to model complex physical processes like mass transfer, \ac{SNe} and \edit1{\ac{CE}} evolution, but one also needs to model large numbers ($10^7$--$10^9$) of systems since many astrophysically interesting phenomena are also rare \edit1{\citep[e.g.,][]{2018MNRAS.481.1908K,2020ApJ...898...71B}}.
Therefore, the computational cost to model each binary must be low to evolve a reasonably sized population with tens of millions of \edit1{binary} systems.
Given a population synthesis code, the number of binaries that need to be simulated to understand an astrophysical population may be reduced by using targeted sampling methods \citep[e.g.,][]{2018ApJS..237....1A,2019MNRAS.490.5228B}; however, there can still be considerable computation cost associated with evolving a binary accurately.
To address this computational challenge fitting formulae were developed to reproduce \ac{sse} across a range of masses and metallicities \citep{Hurley2000MNRAS.315..543H}.
The \ac{sse} equations were then combined with recipes to approximate the evolution of binary stars in the population synthesis code \acl{bse} \citep[\acsu{bse};][]{Hurley2002MNRAS.329..897H}. 
This approach was computationally efficient enough for population studies, but came at the cost of physical approximations (e.g., simple parameterizations for the stability of mass transfer, lack of self-consistency in the treatment of stars out of thermal equilibrium and of stellar wind mass loss).

Since the development of \ac{bse}, similar rapid population synthesis codes have been developed using the same underlying methodology of relying on the \ac{sse} fitting formulae for single star evolution and different recipes to approximate binary evolution. 
These include \edit1{\texttt{SeBa} \citep{1996A&A...309..179P,2012A&A...546A..70T}, \texttt{Scenario Machine} \citep{1996smbs.book.....L,2009ARep...53..915L}}, \texttt{StarTrack} \citep{2002ApJ...572..407B,2008ApJS..174..223B}, \texttt{binary\_c} \citep{2004MNRAS.350..407I,2006A&A...460..565I,2009A&A...508.1359I}, \texttt{COMPAS} \citep{2017NatCo...814906S,2018MNRAS.477.4685B,2022JOSS....7.3838C}, \texttt{MOBSE} \citep{2018MNRAS.474.2959G}, and \texttt{COSMIC} \citep{2020ApJ...898...71B}. 
There are other approaches to population synthesis diverging from using SSE fitting formulae, aiming to increase the accuracy of physics being implemented.
The \texttt{ComBinE} \citep{2018MNRAS.481.1908K} and \texttt{SEVN} \citep{2015MNRAS.451.4086S,2019MNRAS.485..889S} codes use dense grids of single star evolutionary sequences and performs linear interpolation between models. 
\texttt{BPASS} provides more accurate modelling of stellar populations as it uses detailed binary models including their interactions although without simultaneously evolving both stars \citep{2017PASA...34...58E}. 
\edit1{Although stellar evolution codes such as \mesa{} can evolve binary star systems self consistently, few population synthesis codes integrate detailed binary evolution models into their framework.}

\ac{posydon} is a new population synthesis code that uses full detailed binary evolution simulations evolved with \ac{mesa} to perform population synthesis \citep{2022arXiv220205892F}.
\edit1{
The \ac{1D} stellar evolution code \ac{mesa} solves the equations of stellar structure and composition as a function of time, allowing \ac{mesa} to self-consistently evolve two stars and their orbit, which is not done with the aforementioned population synthesis codes.
A few binary effects that are important to model self-consistently include the structural response to mass loss in each star, as well as angular momentum transport between the orbit and stellar rotations.
Modeling binary systems with \ac{mesa} provides greater physical accuracy than binary population synthesis modeling, but at a comparatively high computational cost.}

Although a single \ac{1D} stellar evolution simulation may have a modest cost ranging from tens to hundreds of CPU-hours, it is common to compute thousands to tens of thousands of models due to the intrinsically large number of parameters that describe stellar and binary physics.
\citep[e.g.,][]{2015ApJ...802L...5F, 2015ApJ...807..184F, 2016ApJ...823..102C, 2020A&A...642A.174M, 2021A&A...650A.107M, 2021ApJ...922..110G,2021A&A...649A.114G,2021ApJ...912L..23R}.
Although the aforementioned studies show that \ac{mesa} simulations can be scaled to run large numbers of systems, they are still too computationally expensive to directly model galactic populations.

The \ac{posydon} framework uses large data sets of precomputed binary evolution simulations to evolve a population through various phases of evolution.
To approximate the evolution of a binary system for which no precomputed model exists, interpolation is employed.
To maximize the interpolation accuracy, and the subsequent population synthesis, the detailed binary simulations must cover an adequate range of parameter space to resolve complex behavior.
Since binary stellar evolution has many intrinsic input parameters (e.g., component masses $M_1$, $M_2$, orbital period $P_\mathrm{orb}$, eccentricity $e$, metallicity $Z$), \edit1{\posydon{} requires} orders of magnitude more models compared to the single star approach of previous codes.

A common procedure for building these data sets is to choose the most important parameters in the problem and create a dense regular (rectilinear) grid of simulations.
Then, the density is generally chosen based upon a combination of physical intuition about the problem and manual inspection  \citep[e.g.,][]{1999A&A...350..148W, 2001ApJ...552..664N, 2007A&A...467.1181D, 2015ApJ...807..184F, 2020A&A...642A.174M, 2021A&A...650A.107M}.
However, \ac{posydon} data sets  of binary simulations will have prohibitively large computational cost to produce when we consider expanding the dimensionality of our data sets.
For example, to construct a grid of detailed binary simulations with 10 points in each of 5 dimensions, assuming a typical computation time of 12 hours per simulation, it would cost $1.2$ million CPU-hours to complete.
\edit1{Here we study how to overcome these computational bottlenecks to use detailed simulations in every major phase of binary evolution.}
We propose active learning as a solution for building grids of binary simulations, at a fraction of the cost of current methods of sampling parameter space.

\ac{AL} is the process of \emph{intelligently} sampling points of interest to build a custom training data set for \ac{ML} algorithms \citep{settles.tr09}.
\ac{AL} is useful in situations where labeling instances (e.g., running a simulation, human annotation of data) is expensive or the number of samples to label is large.
Depending on the application, AL algorithms optimize for either classification or regression with applications including speech recognition, image classification, and information extraction \citep{2020Kumar}.
\edit1{In astronomy} it has been proposed as a way to optimize spectroscopic follow-up for labeling supernovae light curves \citep{2019MNRAS.483....2I,2020arXiv201005941K}, as well as mitigating sample selection bias for photometric classification of stars \citep{2012ApJ...744..192R}.
It has also been applied to multiple computational astrophysics problems to build training sets for interpolation or increased performance over standard sampling methods \citep[e.g.,][]{2005MNRAS.363..543S,2017PhRvD..96l3011D,2022PhRvR...4a3046R,2022arXiv220507987D}.
\edit1{Here we apply \ac{AL} to population synthesis.}

In this paper we present an \ac{AL} framework for dynamically constructing data sets of binary simulations to be used in population synthesis codes such as \ac{posydon}.
In \secref{sec:methods} we introduce our new AL algorithm, \psycris{} (PoSYdon Classification and Regression Informed Sampling), and describe our implementation. Then we present two tests of \psycris{}: first on a synthetic data set \secref{sec:synthetic_data_test} and then in a real-world scenario using \ac{mesa} to run binary evolution simulations \secref{sec:mesa_test}. 
In \secref{sec:discussion} we discuss the results of our tests.
We find that \psycris{} performs well when optimized on synthetic data and we see promising performance when applying our method to construct a data set of \ac{mesa} simulations. 
The \psycris{} code is open source, and will be available as a module of \ac{posydon}.
The \psycris{} algorithm is broadly applicable to other AL problems with both classification and regression data.

\section{The \psycris{} Algorithm
\label{sec:methods}} 

    \subsection{Defining the Problem 
    \label{sec:sub:Defining the Problem}}
    
    To describe the \psycris{} algorithm, we first provide a description of the data set and the problem we are trying to solve in the context of binary population synthesis with \posydon{}.

    An individual simulation corresponds to a vector $\Simul{}(t) = [v_{1}(t), \dots, v_{n}(t) ]$ that evolves in time $t$ based on some physical model \PhysModel{}.
    We can write the evolution of a simulation from an initial state $\Simul{i}$ to some final state $\Simul{f}$ with our model \PhysModel{} as $\PhysModel{}(\Simul{i}) = \Simul{f} = [v_{1}(t_f), \dots, v_{n}(t_f) ]$.
    Our task is to predict the outcomes of all simulations in a given simulation domain \Ddata{}, without manually running simulations at all points in the space due to the prohibitive computational cost.

    In the case of \posydon{}, \Ddata{} would consist of parameters describing binary stellar evolution (e.g., component masses $M_1$ and $M_2$, orbital period $P_\mathrm{orb}$, metallicity $Z$, stellar rotation $\omega_1$ and $ \omega_2$, eccentricity $e$).
    Our physical model \PhysModel{} is \mesa{}, evolving a binary system from some initial configuration to a final state.
    Then, $v(t)$ is anything reported by, or post-processed from the \mesa{} simulations, that change over the evolution of a binary. 
    For example, it could be a real number, like the mass of the star at time $t$, or it could be a classification, such as whether a star is a \ac{BH} or a \ac{NS}.
    \edit1{Finally, $\Simul{}(t)$ is the set of all of these output parameters for a single binary.}
    In \posydon{}, we use \ac{ML} techniques to predict the outcomes of these \mesa{} simulations to evolve a binary population through various phases of evolution.

    It is helpful to split our domain into two categories: the set of labeled data \Ldata{}, completed simulations with inputs and outputs, and unlabeled data \Udata{}, the space of possible simulations which have not been computed.
    To evolve simulations $\Simul{i} \in \Udata{}$, we can use a \ac{ML} model \MlModel{} trained on a labeled data set \Ldata{}, denoted as $\MlModel{}_{\Ldata{}}$.
    Then the predicted evolution of a simulation is given by:
    \begin{equation}
        \MlModel{}_{\Ldata{}}( \Simul{i} ) = \Simul{f}^{*},
    \label{eqn:mlmodel_pred_sim}
    \end{equation}
    where the star in $\Simul{f}^*$ denotes predicted final properties.
    For many computational-astrophysics problems, the computational cost of evaluating $\MlModel{}_{\Ldata{}}(\Simul{i})$ is much less than $\PhysModel{}(\Simul{i})$. 
    \edit1{For example, getting a prediction from a \ac{ML} algorithm trained on \mesa{} data is much less expensive than running \mesa{}.}
    To be applicable in any scientific studies, we must also ensure that our \ac{ML} model has high accuracy, i.e., $\Simul{f}^* \approx \Simul{f}$.
    However, we must also consider the cost of computing our training set \Ldata{} and then optimizing \MlModel{}, each of which is not always negligible \edit1{\citep[e.g., for training Gaussian processes;][]{Rasmussen2006}}.

    Our study focuses on identifying an optimal \Ldata{}, without knowing a priori what our simulations across \Ddata{} may look like, at a minimized computational cost.
    Our \ac{AL} algorithm \psycris{} takes an initially sparse training data set $\Ldata{}$, and identifies new points $\Simul{i} \in \Udata{}$ such that when they are labeled \edit1{($\PhysModel{}(\Simul{i}) = \Simul{f}$)} and added to the training set for \MlModel{} \edit1{($\Ldata{} = \Ldata{} \cup \Simul{f}$)}, we achieve high accuracy ($\MlModel{}_{\Ldata{}}(\Simul{i}) \approx \PhysModel{}(\Simul{i})$) with a low computational cost compared to standard methods of constructing the training set.

    \subsection{Algorithm \& Workflow
    \label{sec:sub:Algorithm & Workflow}}
    
    \Acl{AL} algorithms take as input an initial training data set and output query points to be labeled by an oracle \citep{settles.tr09}. 
    The oracle may be a human manually labelling data or the results of simulations that we treat as the truth (as in \posydon{}).
    \Acl{AL} algorithms work in a loop with the oracle to iteratively build a training data set designed to achieve high accuracy in interpolation with an economical number of simulations. 
    It is common for \ac{AL} algorithms to be applied in either classification or regression problems \citep{2020Kumar}, but in \posydon{} our data naturally span both regimes.
    For example, one can classify binary simulations using their mass transfer histories and perform regression on their continuous output quantities like final orbital separation, age, and component masses.
    Therefore, we designed \psycris{} to consider both classification and regression simultaneously.
    
    We achieve this by first combining \ac{AL} \emph{heuristics}, which are estimates for uncertainty in classification or regression (\secref{sec:sub:sub:combining_C_and_R}).
    Next, we sample from this new, combined AL heuristic and generate an \emph{uncertainty distribution} (\secref{sec:sub:sub:sampling_query_points}).
    Finally, we draw query points from the uncertainty distribution in serial or batches.
    This completes a single \ac{AL} loop of \psycris{} which will repeat until a termination condition is met.
    This condition can simply be a maximum number of sampled points, or something more complex.
    The details of the \psycris{} algorithm are described in \autoref{sec:sub:Components & Concepts}, our exact implementation is described in \autoref{sec:sub:Implementation}, and our results from applying \psycris{} are presented in \autoref{sec:synthetic_data_test} and \autoref{sec:mesa_test}.

\begin{figure*}
        \centering
        \includegraphics[width=2\columnwidth]{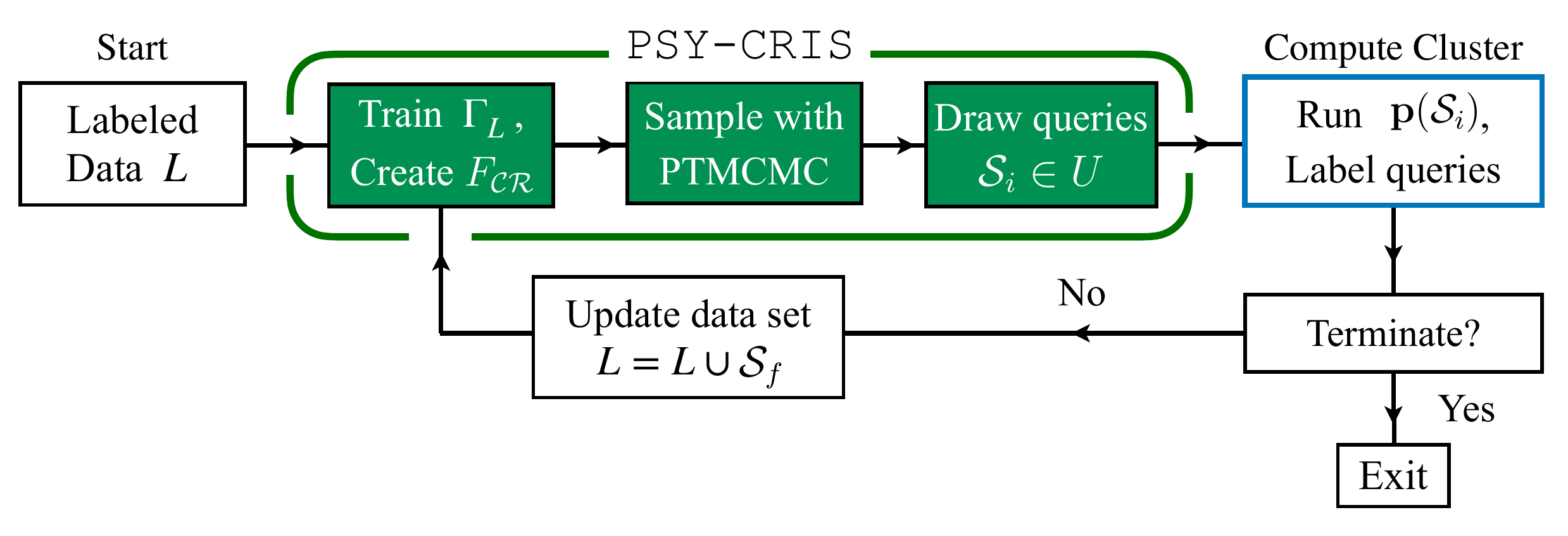}
        \caption{
        A schematic \ac{AL} flowchart showing how \psycris{} integrates with a compute cluster to submit and label \mesa{} simulations using \posydon{} infrastructure. 
        In our case, \PhysModel{} is \mesa{}, but in general it may be switched out for any other simulation software.
        The loop starts from a set of labeled data \Ldata{} given to \psycris{}, which outputs a set of new query points $\Simul{i}$ to be labeled.
        The term $F_\mathcal{CR}$ is our custom \ac{AL} heuristic which combines classification and regression simultaneously (\autoref{eqn:AL_metric_CR}).
        After the queried \mesa{} simulations complete, \posydon{} post processes and labels the outcomes of each simulation (\Simul{f}).
        Then, the \ac{AL} loop can continue by feeding the updated training set back into \psycris{}, or it can terminate and exit.
        The final result after multiple iterations is a training data set \Ldata{AL} designed to provide better performance for classification and regression algorithms compared to a simpler distribution of training data.
        }
        \label{fig:schematic_psycris_diagram}
    \end{figure*}

    \autoref{fig:schematic_psycris_diagram} shows a schematic diagram of \psycris{} integrated with \posydon{} to run binary stellar evolution simulations with \mesa{}.
    The green rectangle shows the scope of the \psycris{} algorithm and how it integrates with other portions of \posydon{} infrastructure.
    \posydon{} runs binary simulations with \mesa{} queried by \psycris{} and then post-processes the output into our pre-defined classes and regression quantities.
    The blue square represents the oracle in an \ac{AL} loop and can be replaced with other simulation software.
    
    \subsection{Components \& Concepts
    \label{sec:sub:Components & Concepts}}
    
    \subsubsection{Classification \& Regression AL Heuristics}
    
    Classification deals with categorical data while regression is used in cases where data are continuously valued.
    For example, we may classify a binary evolution simulation based on its mass transfer history while a regression quantity may be the final mass of each component.
    Conceptually, a classification or regression AL heuristic is designed to locate unlabeled samples of interest based on a given \ac{ML} method applied to \Ldata.
    If the \ac{ML} algorithm has a measure of uncertainty in its predictions, or if one can be constructed, then this can be used as an \ac{AL} heuristic.
    We denote a general classification or regression \ac{AL} heuristics as $\mathcal{C}$ and $\mathcal{R}$ respectively.

    \subsubsection{Combining Classification \& Regression AL Heuristics \label{sec:sub:sub:combining_C_and_R}}
    
    We designed the \psycris{} algorithm to optimize for both classification and regression simultaneously.
    To enable this coupled optimization, we propose that a linear combination of individual heuristics can also be used as a heuristic.
    We denote our new heuristic as $F_{\mathcal{CR}}$ defined as:
    \begin{equation}
        F_{\mathcal{CR}}(\mathbf{x}) = 
        \left[ 
        \tau F_{\mathcal{C}}(\mathbf{x}) + (1-\tau) F_{\mathcal{R}} (\mathbf{x})
        \right]^{\beta}
        \label{eqn:AL_metric_CR}
    \end{equation}
    where $\mathbf{x}$ is a point in parameter space, $\tau \in [0,1]$ sets the fractional contribution of the classification and regression terms, and $\beta$ controls the sharpness of the distribution \citep[similarly to an inverse temperature;][]{2018arXiv180508916M}.
    The terms $F_{\mathcal{C}}$ and $F_{\mathcal{R}}$ are normalized heuristics.
    Since we combine $\mathcal{C}$ and $\mathcal{R}$ terms, their relative scaling matters.\footnote{We can normalize a numerically unbounded \ac{AL} heuristic by using an activation function like the logistic sigmoid.}
    \autoref{eqn:AL_metric_CR} is designed to be high valued in regions with the largest estimated uncertainty in classification and regression.
    \edit1{This is just one way to combine $\mathcal{C}$ and $\mathcal{R}$, but many other choices could be made.}

    \subsubsection{Sampling Query Points \label{sec:sub:sub:sampling_query_points}}
    
    Once the heuristic has been defined, we must select query points for the oracle to label.
    \Acl{AL} algorithms can return either a single query point (serial) or a set of query points (batch) to label. 
    Batch proposal schemes introduce more complexity, but allow multiple queries to be labeled in parallel. 
    This may be less efficient in terms of the total number of simulations needed than only ever picking the best point, but provides a advantage in terms of computational wall time \citep{2000Schohn,2017Cai,2017arXiv170800489S,2020arXiv201005941K}.
    With batch proposals, we \edit1{may also} consider more than just the maximum of the \ac{AL} heuristic, but also the diversity of samples in the batch to avoid redundancy \citep{2004Natur.427..247K,2006Hoi,2011ITGRS..49.1014D,2017ITGRS..55.3071W}.

    In \psycris{} we adopt a flexible approach which allows either serial or batch proposals, by creating an \emph{uncertainty distribution} from which any number of query points can be drawn.
    We use our combined heuristic in \autoref{eqn:AL_metric_CR} as the target distribution for a \ac{PTMCMC} sampling algorithm \citep{1986PhRvL..57.2607S}.
    \edit1{\Acl{PTMCMC} \citep{2005PCCP....7.3910E} combines multiple Markov-chain Monte Carlo sampling chains at different \emph{temperatures} (where the coldest  temperature corresponds to the target distribution, and an infinite temperature a uniform distribution), to improve the exploration of parameter space, which can be challenging for standard Markov-chain Monte Carlo if the target distribution is sharply peaked and or multimodal.}
    We assume that \psycris{} may query any point in the input space, and is not constrained to a fixed or discrete pool of unlabeled samples.
    Drawing query points is the final step in one \psycris{} iteration.

    \subsection{Implementation
    \label{sec:sub:Implementation}}
    
    \subsubsection{Classifiers \& Regressors}
    
    We perform one-against-all binary classification. 
    For data with $n$ classes (where $n>2$), we train $n$ binary classifiers and define the predicted class $\hat y$ and classification probability $P( \hat y | \mathbf{x})$ respectively
    \begin{equation}
        \hat y = \argmax\limits_{i=1,\ldots,n} [\theta_i(\mathbf{x})],
    \label{eqn:predicted class}
    \end{equation}
    \begin{equation}
        P( \hat y | \mathbf{x}) = \max\limits_{i=1,\ldots,n} [\theta_i(\mathbf{x})],
    \label{eqn:classification probability}
    \end{equation}
    where $\mathbf{x}$ is a query point in parameter space and $\theta_i$ denotes a binary classifier for the $i$-th class.
    The classifier $\theta_i(\mathbf{x})$, returns the probability that a query point $\mathbf{x}$ corresponds to class $i$.

    To construct our classification heuristic we use a \emph{least confident} measure for classification \citep{1994cmp.lg....7020L,settles.tr09,2020Kumar}:
    \begin{equation}
        F_\mathcal{C}(\mathbf{x}) = 
        \frac{n}{n - 1}\left(1 - P( \hat{y} | \mathbf{x}) 
        \right) ,
    \label{eqn:TD_classifcation}
    \end{equation}
    where $\mathbf{x}$ is a query point, $n$ is the total number of classes in the data set, and $P( \hat y | \mathbf{x}) $ is the classification probability in \autoref{eqn:classification probability}.
    The prefactor is a pseudo-normalization term in the event all $n$ classes intersect at some point in the domain, which need not occur.
    
    We train regression algorithms on data separated by class, taking into account both the different numbers of outputs and unique outputs per class. 
    For $n$ classes that all have $m$ outputs, there are $n \times m$ interpolation algorithms trained.

    For \ac{AL} regression problems, approaches for estimating uncertainty include minimizing estimates of model variance or training multiple models to identify areas of disagreement.
    In \psycris{}, we use a simple and computationally inexpensive measure: the average difference in the output between $3$ nearest neighbors in the training set within the same class.
    We use the standard Euclidean metric to find nearest neighbors in parameter space.
    This is essentially a probe of the local function change in a given class.
    We calculate the average distance between $N$ nearest neighbors in the $i$-th class as follows:
    \begin{equation}
        g_j^i( \mathbf{x} ) = \frac{1}{N} \sum^{N}_{k=1} | f_j^i(\mathbf{x}) - f_j^i(\mathbf{x}_k) | ,
    \label{eqn:regression_abs_diff} 
    \end{equation}
    where $\mathbf{x} \in \Ldata{}$ is a point in the labeled training set, $f_j^i(\mathbf{x})$ is the regression output in the $j$-th output variable in that class, and $k$ iterates through the $N$ nearest neighbors in the class.
    Then we calculate \autoref{eqn:regression_abs_diff} for all points in the training set \Ldata{}.
    
    Since \autoref{eqn:regression_abs_diff} can only be calculated at points in the training set, we interpolate between values of $g_m^i$ to give a continuous distribution for any unlabeled query point $\mathbf{x} \in \Udata{}$. 
    This will be utilized during sampling where we must calculate \autoref{eqn:AL_metric_CR} at any point in parameter space.
    
    In order to combine \autoref{eqn:regression_abs_diff} with a classification term, we pseudo-normalize it so that both terms are of the same order:
    \begin{equation}
        F_\mathcal{R}(\mathbf{x}) = 
        A_1 \log_{10}\left[ A_0 \max\limits_{j=1,...m}[ g_{j}^i (\mathbf{x}) ] + 1 \right].
    \label{eqn:TD_regression}
    \end{equation}
    The constant $A_0$ sets the scale of important absolute differences in the data set, while $A_1$ sets the scale of the function itself.
    We set $A_0 = 0.5$ and use $A_1$ as a normalization term, calculated by inverting the maximum value possible for \autoref{eqn:TD_regression} across all classes, which is specific to the training data set.
    We take the maximum absolute difference across multiple regression outputs in the event that a class contains more than one.
    Numerically, \autoref{eqn:TD_regression} is monotonically increasing, similar in form to the softplus activation function.
    
    Our choices for $F_\mathcal{C}$ and $F_\mathcal{R}$ in our implementation of \psycris{} are motivated by simplicity and are not to be taken as standard or optimal.
    For a more complete review of \ac{AL} heuristics, see \citet{settles.tr09}. 

    \subsubsection{Sampling Method}
    
    After defining $F_\mathcal{C}$ and $F_\mathcal{R}$, we combine the two terms using \autoref{eqn:AL_metric_CR} to use as the target distribution for a \ac{PTMCMC} sampling algorithm.
    Our implementation of \ac{PTMCMC} uses one walker for each temperature in the chain, where each walker's temperature is given by the relation $T_{i+1} = T_{i}^{1/c}$, where the spacing constant $c=1.2$.
    The number of chains is then defined from the maximum temperature $T_\mathrm{max}$ down to $T=1$.
    We use the Metropolis--Hastings jump proposal scheme and take three steps before chain swap proposals \citep[][]{1953JChPh..21.1087M, 10.1093/biomet/57.1.97}. 
    The stochastic sampling of points may help with exploring the parameter space, and discovering new regions with class boundaries \citep{2009Joshi,2018Yang}.
    
    The \psycris{} algorithm generalizes for both serial and batch proposal schemes since any number of points can be drawn directly from the uncertainty distribution.
    Although this approach does not guarantee optimally spaced points within a batch, more complicated algorithms can be easily adopted into our formalism.

    \subsubsection{Handling small classes}
    
    It is possible for the \psycris{} algorithm to discover classes by random chance, that were not in the original training set.
    Initially, new classes may contain only one sample.
    Our one-against-all binary classification scheme can handle this, but our regression algorithms are separately trained by class.
    Therefore, fitting most regression algorithms for a small class (one data point) is not well defined and fails.
    We could set the regression term $F_\mathcal{R}$ to zero in \autoref{eqn:AL_metric_CR} if small classes are unimportant, but in our application, many astronomically interesting systems are also rare and may be subject to this edge case.
    
    Therefore, to reflect the importance of small classes (and to avoid the numerical limitation of fitting small classes), we implement a separate proposal technique which supersedes the regular \psycris{} algorithm in the event a small class is present in the training set.
    We draw points from a multivariate normal centered on the small class (one point) with a length scale set by the $10$ nearest neighbours in input space (regardless of class) and scaled arbitrarily by $1/30$ which was found to work well in tests.
    Then if the first proposal iteration does not populate the small class, subsequent iterations will tend to decrease the length scale.
    \edit1{In practice, we would take care to construct our initial training set to contain an adequate sample of all relevant classes.
    The procedure described here is only necessary to test \psycris{} in non-ideal conditions.}

\subsection{Performance Metrics
\label{sec:sub:performance_metrics}} 

To evaluate the performance of our algorithm, we compare training sets built with \psycris{} to baseline configurations including regularly spaced grids and random uniformly distributed data.
We use a validation set with known labels to determine how well classification and regression algorithms extrapolate after being trained on the different training sets.

For classification we calculate the overall classification accuracy \ClassAcc{}, which we define as 
\begin{equation}
    \ClassAcc{} = \frac{ N_{\mathrm{pred,\mathbf{C}}} }{ N_{\mathrm{total}} },
    \label{eqn:class_accuracy}
\end{equation}
where $N_{\mathrm{pred,\mathbf{C}}}$ is the number of correct predictions and $N_{\mathrm{total}}$ is the total number of queries.

We also use the per-class classification accuracy which we define as 
\begin{equation}
    \perClassAcc{} = \frac{ N^\mathrm{i}_{\mathrm{pred,\mathbf{C}}} }{ N^\mathrm{i}_{\mathrm{total}} },    \label{eqn:per_class_accuracy}
\end{equation}
where $N^\mathrm{i}_{\mathrm{pred,\mathbf{C}}}$ is the total number of correct predictions for the $i$-th class and $N^\mathrm{i}_{\mathrm{total}}$ is the total number of points that truly belong to that class (true positives and false negatives).
\edit1{Our definition for \perClassAcc{} is consistent with what is often referred to as \emph{recall} of a classifier}.
Although similar to \ClassAcc{}, \perClassAcc{} can trivially approach $1$ by over-predicting the classification region which often happens for small training sets: \perClassAcc{} is sensitive to false negatives but insensitive to false positives.

\edit1{We also calculate the F1-score, the harmonic mean of recall and precision, to compare with \autoref{eqn:per_class_accuracy}, and find both \perClassAcc{} and the F1-score have qualitatively similar behavior for the tests we perform.}

For regression accuracy (\RegrAcc{}) we use the $90$-th percentile of the distribution of absolute differences between interpolation predictions and true values from the validation data set.
In \psycris{} we train regression algorithms on data organized by class, so differences calculated at any point must first be classified to select the appropriate interpolator.
Therefore, errors in classification can affect \RegrAcc{} by requiring predictions in regression for points well outside the true classification region.
We could consider calculating absolute differences only for true positive classifications, but in practice we will not know the true outputs at every point.
We choose to calculate absolute errors for points predicted to be in a given class, and only remove data if a class has no true regression data to compare against the prediction.
We calculate a combined \RegrAcc{}, where all absolute differences are combined in to a single distribution regardless of class, and a per-class \RegrAcc{}, where each class has its own absolute difference distribution.

For our regression algorithm, we use \ac{RBF} interpolation, which is parameterized by a length scale.
By default the length scale is set to the average distance between training data which is assumed to be a good starting point.
For the synthetic data test \secref{sec:synthetic_data_test}, we set the length scale manually to the wavelength from \autoref{eqn:analytic-regression}, as it was found to achieve higher accuracies in small tests.
For the \mesa{} test \secref{sec:mesa_test}, we allow the default behavior, instead of explicitly attempting to fit for an optimal length scale.
\section{Synthetic Data Test}
\label{sec:synthetic_data_test}

\subsection{Test Setup}
\label{sec:sub:synth_data_setup}
To determine how \psycris{} performs compared to standard methods of constructing training data sets, we first perform a test applying \psycris{} on synthetic data.
The use of synthetic data with known underlying distributions allows us to test the performance of the algorithm without resolution concerns.
We construct a \ac{3D} synthetic data set including both classification and regression data drawn from analytic functions.
The data contain six unique classes and one regression output which is continuous across class boundaries.
A detailed description of the synthetic data set, including the analytic functions and a visualization of the classification space, can be found in \autoref{sec:appendix}.

Using the aforementioned synthetic data set, we start every \ac{AL} run from a sparse, regularly spaced $5\times5\times5$ grid.
We let \psycris{} iteratively query new points, which are labeled by the oracle in the standard \ac{AL} loop, until reaching a combined total of $10,000$ points in the final labeled data set.
In addition to testing a default configuration of \psycris{}, we run a suite of different models varying the contribution factor $\tau$ and the sharpness parameter $\beta$ (as in \autoref{eqn:AL_metric_CR}), to demonstrate their effect on our results.
When evaluating performance we compare our resulting training set from \psycris{} (\Ldata{AL}) to randomly distributed training data (\Ldata{RS}) as a benchmark and regularly spaced training data (\Ldata{Grid}).
\edit1{The test set for each configuration of training data is an over-dense regular grid with $10^6$ points.}

\begin{figure*}
    \centering
    \includegraphics[width=2\columnwidth]{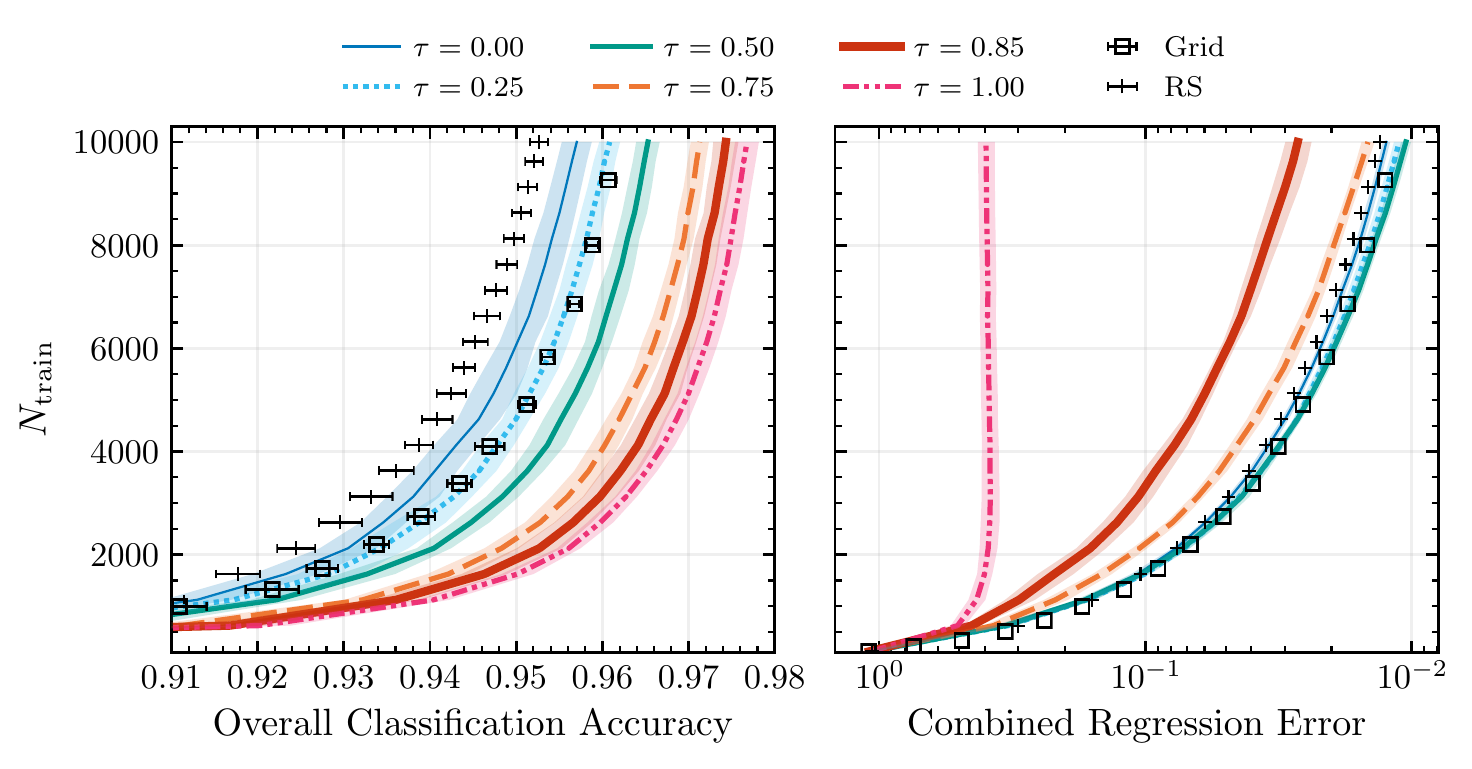}
    \caption{The number of training points needed to achieve a given classification (left panel) and regression (right panel) performance in our synthetic data set test.
    We define overall classification accuracy using \autoref{eqn:class_accuracy} and define regression accuracy as the $90$-th percentile of the distribution of absolute differences.
    We compare our \ac{AL} algorithm \psycris{} (solid lines) to regularly spaced grids (Grid), and random sampling (RS) while varying $\tau$ and fixing $\beta=2$.
    In both panels, high performance data sets exist in the bottom right corner, reaching high accuracy or low error with a small training set.
    Error bars in the combined regression error for Grid and RS are comparable to the size of their respective markers.
    The \psycris{} configuration with $\tau=0.5$ outperforms the regular grid and random sampling in both classification and regression simultaneously. 
    }
    \label{fig:absdiff_overall_cls_and_regr}
\end{figure*}

\begin{figure*}
    \centering
    \includegraphics[width=2\columnwidth]{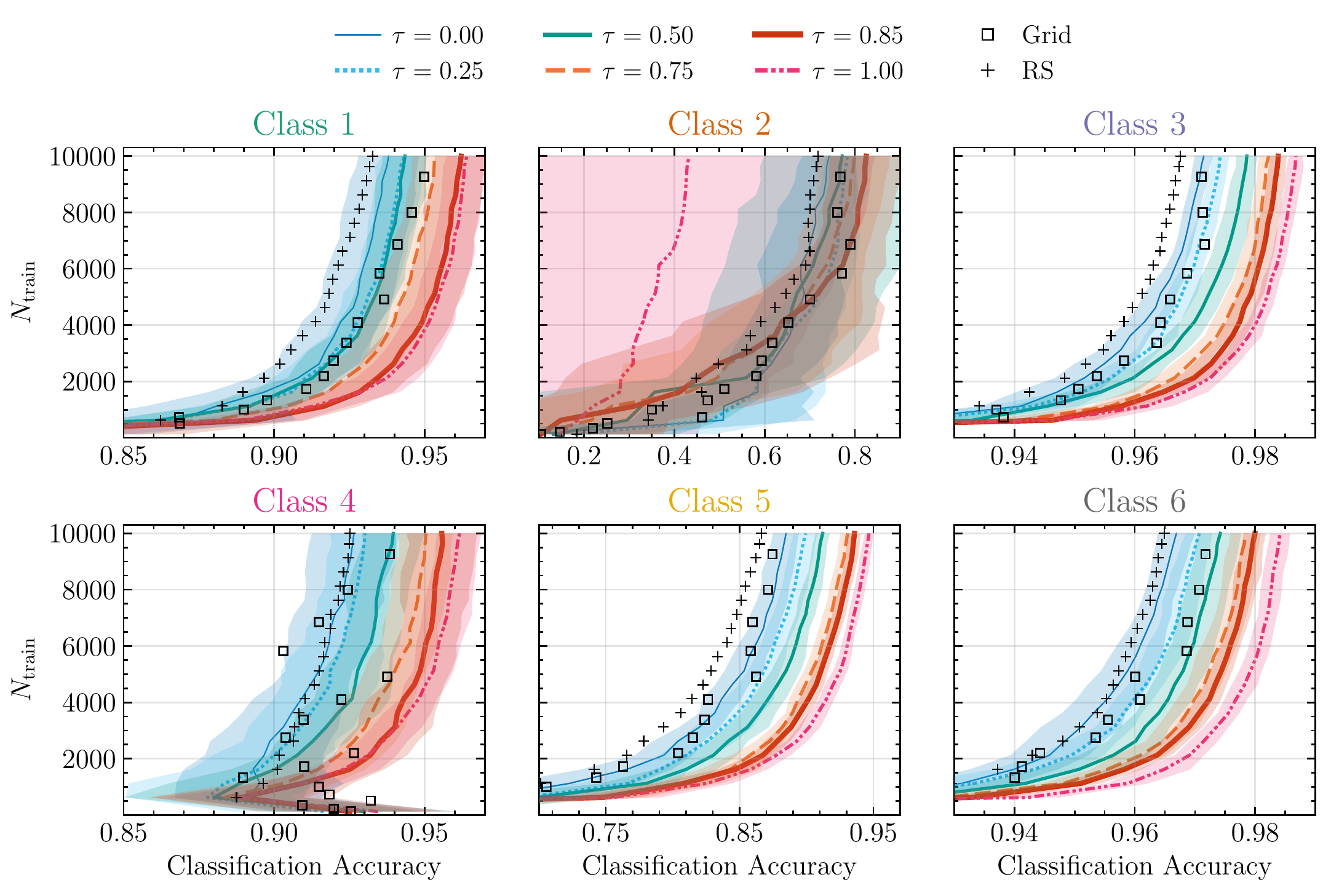}
    \caption{The number of training points needed to achieve a given classification accuracy in our synthetic data test, separated by class.
    All lines and markers are the same as in Figure~\ref{fig:absdiff_overall_cls_and_regr}.
    We find that all classes exhibit qualitatively similar behavior in performance and improvements from AL.
    However, features such as the average performance at large $N_{\rm train}$ and the improvements between various \ac{AL} configurations (different $\tau$), are unique between classes.
    Class 2 exhibits the largest variance and lowest average performance due to its small size and the fact that it must be discovered by \psycris{} in some cases as it may not be present in the starting training set.
    Class 5 sees the largest gains in performance from \ac{AL} because of its extended, narrow shape in parameter space (\autoref{fig:synth_data_3D}) which \psycris{} is able to resolve.
    The errors for random sampling points are similar to the errors shown for the $\tau=0$ line for all classes.
    The errors for grid points are generally half that width except for Class 2 which has comparable errors throughout.}
    \label{fig:per_class_classification}
\end{figure*}

\begin{figure*}
    \centering
    \includegraphics[width=2\columnwidth]{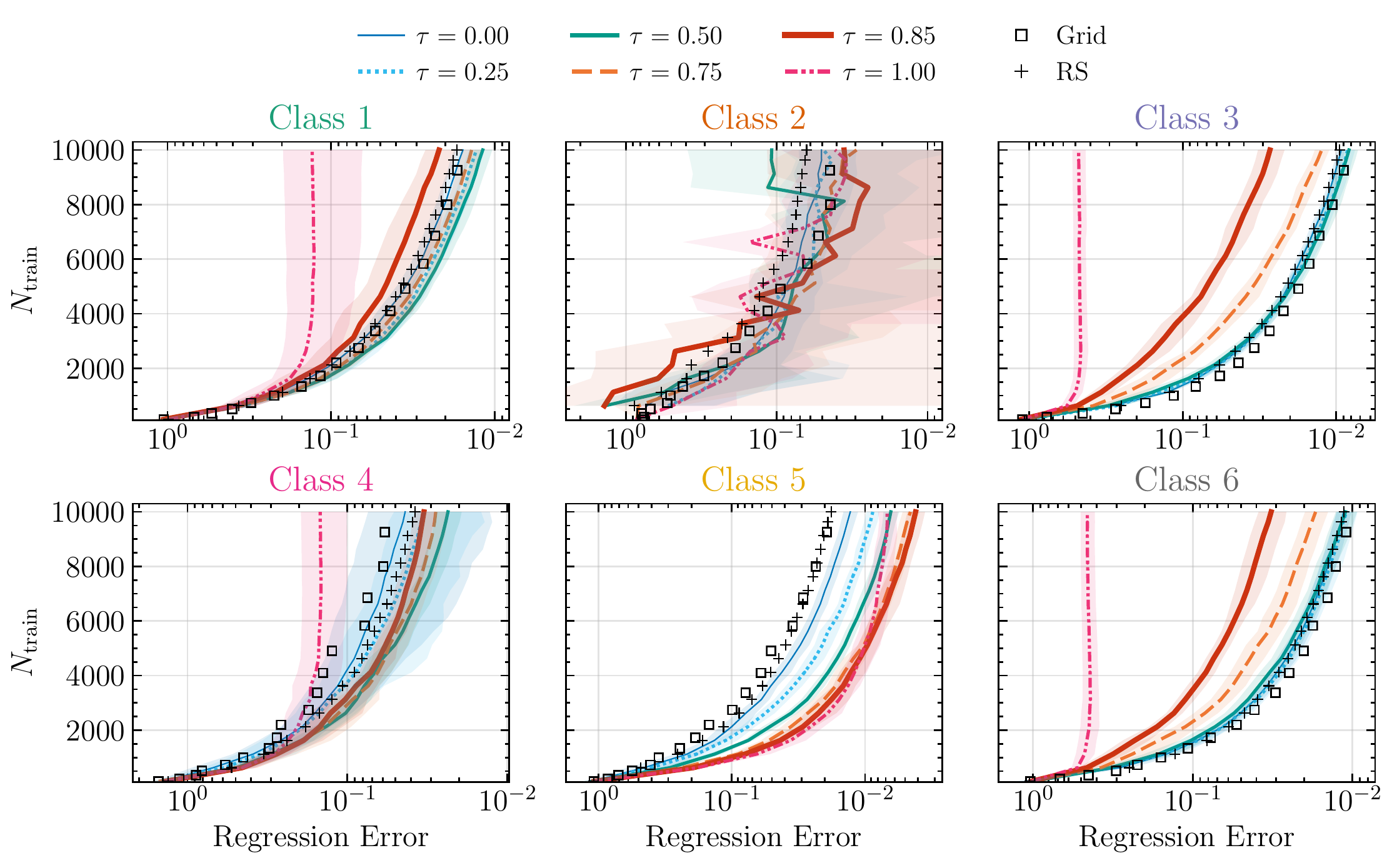}
    \caption{The total number of training points needed to achieve a given regression performance (defined as the $90$-th percentile of the distribution of absolute differences) in our synthetic data set, separated by class. 
    We compare our \ac{AL} algorithm \psycris{} (solid lines) to regularly spaced grids (Grid), and random sampling (RS) while varying $\tau$ and fixing $\beta=2$.
    We find that for some classes, our best performing configurations of \psycris{} slightly outperform Grid and RS.
    In most cases, the best performing configuration of \psycris{} is not $\tau=0$ (which we might expect after seeing the trends in \autoref{fig:per_class_classification}), but an intermediate value of $\tau$ which itself is not consistent across classes.
    Class 5 shows significant gains in regression performance compared to Grid and random sampling, but with the best performing \psycris{} configurations focusing more on classification ($\tau>0.5$).
    However, this behavior is reasonable in cases where classification is challenging, and we see Class 5 has the greatest increases in classification performance when using \ac{AL} compared to random sampling and regular grids (\autoref{fig:per_class_classification}).
    The errors for grid points are comparable to the size of the markers for all classes.
    The errors for random sampling points are similarly small except for Class 2 and Class 4 which exhibit errors on the order of the $\tau=0$ line.}
    \label{fig:ad_per_class_regr}
\end{figure*}

\begin{figure*}
    \centering
    \includegraphics[width=2\columnwidth]{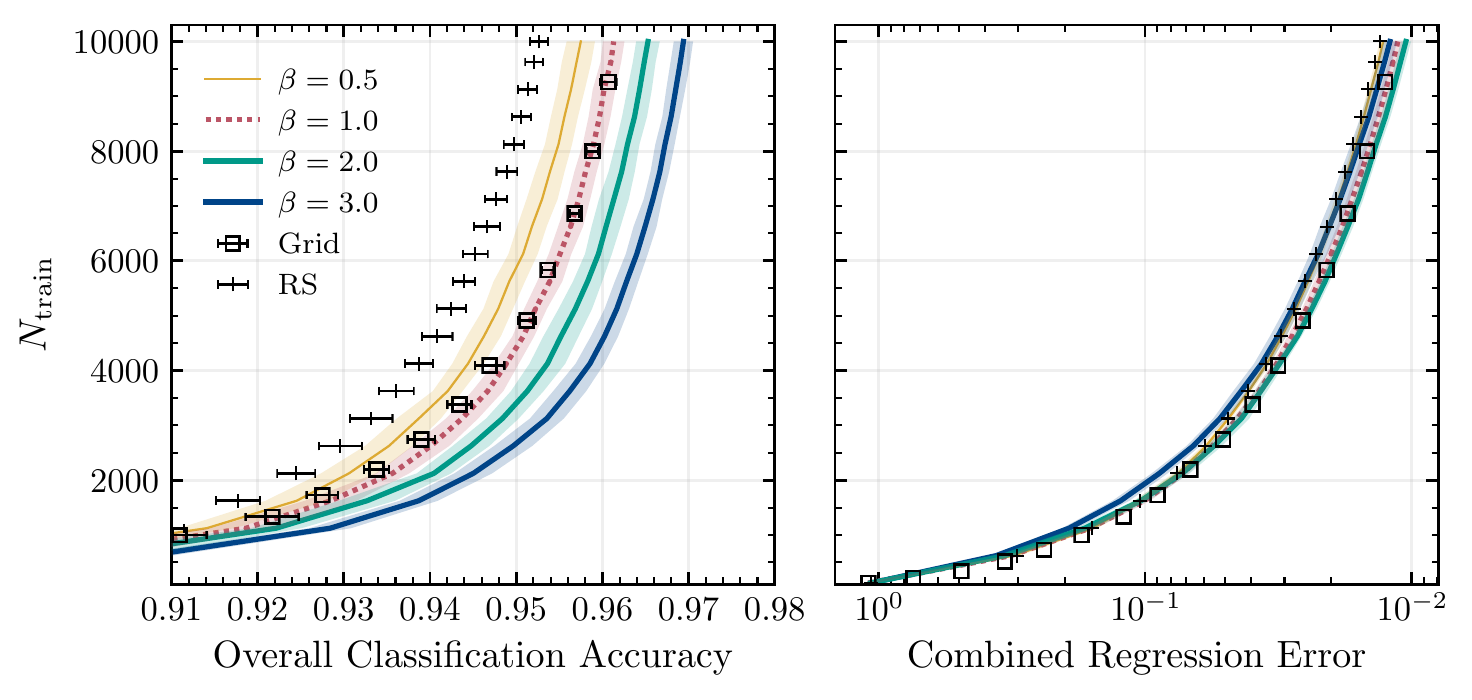}
    \caption{The same as \autoref{fig:absdiff_overall_cls_and_regr} but we fix $\tau=0.5$, the contribution between $\mathcal{C}$ and $\mathcal{R}$ terms, and vary the sharpening parameter $\beta$ as in \autoref{eqn:AL_metric_CR}.
    Increasing $\beta$ results in a higher classification accuracy while not significantly affecting the regression performance.
    However, at $\beta=3$ we see the regression performance drop slightly.
    The right choice of $\beta$ will depend on the data set as well as the classification and regression algorithms implemented.
    }
    \label{fig:absdiff_overall_cls_and_regr_beta}
\end{figure*}

\subsection{Synthetic Data Test Summary}
\label{sec:sub:synth_data_summary}

\autoref{fig:absdiff_overall_cls_and_regr} shows aggregate performance metrics comparing AL to random sampling and regularly spaced grids at various densities.
Since the \psycris{} algorithm contains inherent variability from \ac{PTMCMC} and query point proposals, we show the average of $15$ \psycris{} runs where the solid lines are the mean and shaded areas show $68$-th percentile 
contours.
To simulate unique \psycris{} runs starting from the same starting grid density, we vary the center of the starting grid randomly by $\pm 1/2$ the bin width in each dimension.
This procedure is also used to determine the variance for regular grid configurations at different densities.

For overall classification accuracy (left panel of \autoref{fig:absdiff_overall_cls_and_regr}) we find that \psycris{} performs significantly better than random sampling (RS) and regularly spaced grids (Grid) when $\tau=1$.
As $\tau$ approaches $1$, we find that each subsequent \psycris{} configuration achieves higher classification accuracies than the last (on the order of $\sim0.5\%$) or reaches a target accuracy with significantly smaller number of training points (e.g., for a $\ClassAcc{}=96\%$, a factor of two reduction in $N_\mathrm{train}$ between Grid and the $\tau=0.75$ model).

For combined regression error (right panel), we see the opposite trend: $\tau$ approaching $0$ gives optimal regression performance. 
However, the best performing \psycris{} configurations do not significantly outperform Grid or RS in regression, and most models converge after $\tau=0.75$.
This large scale behavior when changing $\tau$ reflects our design principles when constructing the target distribution in \autoref{eqn:AL_metric_CR}.

When considering classification and regression simultaneously with $\tau=0.5$, \psycris{} outperforms Grid and RS in classification, \emph{and} performs similarly to Grid and RS in regression.
Conversely, considering only classification ($\tau=1$) or regression ($\tau=0$) alone during \ac{AL} may lead to significant losses in performance for the neglected category.

We may also consider performance on a class-by-class basis, allowing a more granular perspective on \ac{AL} performance across the domain.
In \autoref{fig:per_class_classification} we show the per-class classification accuracy defined in \autoref{eqn:per_class_accuracy}.
We find the trends between models at different $\tau$ remain similar to those in overall classification, but some classes achieve larger (Class 5, $\tau=0.5$) or smaller (Class 1, $\tau=0.5$) gains from \ac{AL}, which is not apparent from \autoref{fig:absdiff_overall_cls_and_regr}.
Class 5 presents a challenging classification problem because it contains an extended narrow region for $z>0$ (\autoref{sec:appendix}), which explains the significant gains from \ac{AL} compared to other classes.
Class 2 is a small class in parameter space that does not necessarily exist in the initial training sets for AL, so it must be discovered by \psycris{} in some cases.
The process of discovering new classes is stochastic, which explains why Class 2 sees the lowest performance compared to all classes at a given $N_\mathrm{train}$.
Class 2 is an extreme example of the ML challenges faced when the training set is highly imbalanced.
In application of \ac{AL}, one must carefully evaluate the problem in question to characterize the expected classes and their scale in the domain such that they are represented in the initial training set \citep{2007Ertekin,2013Attenberg}.

In \autoref{fig:ad_per_class_regr} we show the per-class regression performance defined as the $90$-th percentile of the distribution of absolute errors.
We find that most classes exhibit trends similar to those seen in the overall regression performance.
For example, Class 1, Class 3 and Class 6 show the best performing \psycris{} runs have $\tau \leq 0.5$.
A clear exception is in Class 5, which achieves the highest performance, exceeding Grid and RS, when the contribution factor $\tau>0.5$ weighs classification more heavily in the target distribution. 
Although this trend may appear to go against our design principles for \psycris{}, we expect classification to impact regression performance in challenging cases because our regression algorithms are trained on data separated by class.
This difficulty in classification is confirmed by the per-class classification performance in \autoref{fig:per_class_classification} where we see Class 5 benefits the most from AL compared to all other classes.
However, considering only classification ($\tau=1$) is not sufficient for reaching the highest regression performance for Class 5 with $\tau=0.85$. 
In summary, \psycris{} performs similarly to regularly spaced grids and random sampling even in the worst cases (e.g., Class 3 and Class 6) and can significantly outperform in others (e.g., Class 5).

We interpret the fact that Grid and RS perform similarly in regression performance to mean that our data can be fit easily with uniformly distributed training data.
This is sensible given the regression function (\autoref{eqn:analytic-regression}) is highly symmetric and smooth, which is not typical for real data.

We also ran models varying the sharpening parameter $\beta$ while fixing $\tau=0.5$, and present aggregate performance results in \autoref{fig:absdiff_overall_cls_and_regr_beta}.
We find that $\beta$ has a stronger effect on the classification performance than regression performance, and that a larger $\beta$ leads to higher classification accuracy.
Although it is difficult to see from \autoref{fig:absdiff_overall_cls_and_regr_beta}, there appears to be a difference of $\sim 1000$ points in $N_\mathrm{train}$ for a constant regression performance from $\beta=2$ to $\beta=3$.
This suggests that $\beta=2$ is close to the optimal value for this data set, and is likely to be a good starting place for problems with similar complexity.

\section{Building A \mesa{} Grid}
\label{sec:mesa_test}

\subsection{\mesa{} test setup}

The primary goal for our \ac{AL} algorithm is to construct an optimal set of detailed binary evolution simulations for use in population synthesis codes, such as the \mesa{} simulations being used in \posydon{}.
An optimal training set provides a target classification and regression accuracy at the lowest computational cost (lowest number of simulations) possible.
In this test we demonstrate \psycris{} working with \posydon{} infrastructure to propose and label \mesa{} simulations (as in \autoref{fig:schematic_psycris_diagram}).
We use a \texttt{python} implementation of Message Passing Interface (MPI), combining the software used to run \mesa{} in \posydon{} with \psycris{}.
Implementing distributed computing is critical as it acts as the oracle in the \ac{AL} loop, automatically running new \mesa{} simulations and post-processing their output.

We model binary systems with a helium (He) main-sequence star and \ac{BH} companion using \mesa{}'s \texttt{binary} module.
We use \posydon{} default \mesa{} controls and input parameters, which are described in detail in the \posydon{} instrument paper \citep{2022arXiv220205892F} with an additional change.
We set the maximum radiative opacity to $0.5~\mathrm{cm}^2\, \mathrm{g}^{-1}$ for all \mesa{} simulations to alleviate numerical convergence issues as described in \citet[][section~6.1]{2022arXiv220205892F}.
Our parameter space is the same as the grid of helium-rich stars with \edit1{\ac{CO}} companions from \citet{2022arXiv220205892F}, which covers the initial masses $M_1$ of $[0.5 M_\odot,80 M_\odot]$, $M_2$ of $[ 1 M_\odot,35.88 M_\odot]$, and the initial orbital period $P_\mathrm{orb}$ of $[0.02~\mathrm{day},1117.2~\mathrm{day}]$.

Part of the \posydon{} infrastructure includes parsing \mesa{} outputs to categorize simulations into one of five classes, used to organize data for interpolation \citep{2022arXiv220205892F}.
The classes are based upon a binary's characteristic evolution: \texttt{initial\_MT} (Roche-lobe overflow at zero-age main-sequence), \texttt{stable\_MT} (dynamically stable Roche-lobe overflow mass transfer), \texttt{unstable\_MT} (dynamically unstable Roche-lobe overflow mass transfer), \texttt{no\_MT} (no Roche-lobe overflow mass transfer), and \texttt{not\_converged} (numerical error or exceeded maximum computation time of two days).
\edit1{Some systems labeled as \texttt{unstable\_MT} can start their mass-transfer sequences stably but eventually evolve to become unstable. 
Any binary sequence labeled as \texttt{stable\_MT} remains stable throughout the entire evolution.}
Numerical convergence issues arise in \mesa{} due to limitations in modeling and often occur in specific regions in parameter space.
In the \ac{AL} phase, we use all five interpolation classes (including the \texttt{not\_converged} systems) and only consider regression for the final orbital period for the \texttt{stable\_MT} and \texttt{unstable\_MT} classes.
The \texttt{not\_converged} class was included to keep \psycris{} from continuously proposing simulations in these problematic regions since any failed run would not add information to the training set, wasting computing resources.
We also log-normalize the inputs and outputs from $[-1,1]$ before entering \psycris{} and transform the proposal points back into linear space before evolving the system with \mesa{}.

Our parameter space for this test is \ac{3D} but can be easily extended into a higher dimensional problem, including covering multiple metallicities to give one example.
\edit1{The design of \psycris{} is generalized to higher dimensions and can readily be applied in such cases.}
We compute a high density regular grid with dimensions $41\times21\times49$ in $\log_{10} \left( M_1 / M_\odot \right)$, $\log_{10} \left( M_2 / M_\odot \right)$, $\log_{10} \left( P_\mathrm{orb}/\mathrm{day} \right)$ respectively totaling $42{,}113$ successful \mesa{} simulations \edit1{($76$ were unable to be parsed due to corrupt output files)}.
We choose these dimensions for the regular grid by inspection based on previous tests in the same parameter space.
We start \psycris{} from a subset of the full density grid by taking every other point in $M_1$ and $M_2$ and every fourth point in $P_\mathrm{orb}$, retaining end points such that the range in each axis matches the full grid.
\edit1{We chose a different sampling for the orbital period because this dimension has a higher resolution because of the large classification changes that occur with small changes in $P_\mathrm{orb}$.}
This procedure creates the starting, low density grid with $3{,}003$ \mesa{} simulations. 
To create the medium-density grid (which we use to compare performance in \secref{sec:sub:mesa_test_summary}) we again use a subset of the full density regular grid,  taking every other point in $P_\mathrm{orb}$, creating a grid with $21{,}525$ \mesa{} simulations.

During the beginning of \ac{AL}, the parent MPI process runs one \psycris{} iteration which is used to propose the initial parameters for all child processes waiting to run \mesa{}.
Once any child process finishes running \mesa{}, the parent process runs \psycris{} again with the updated data set and proposes a new point in parameter space to run with \mesa{} (\autoref{fig:schematic_psycris_diagram}).
This process is effectively a serial proposal scheme after the initial startup phase, although it depends on the rate at which \mesa{} simulations finish. 
In our test, \psycris{} terminates when a maximum number of proposed points is reached.
We stop this test when we could see trends in our performance metrics since this demonstration is computationally expensive.
In practice \psycris{} is meant to be used for grid sampling until a threshold accuracy in classification, regression, or a combination of the two is achieved.
We use a fiducial configuration of \psycris{} with parameters taken from the synthetic data tests with $\tau=0.5$ and $\beta=2$.

Unlike with the synthetic data set, the exact outcomes of all \mesa{} simulations in our simulation domain are not known exactly.
In order to calculate our standard performance metrics, we create a \edit1{test} set by running $10,000$ random uniformly distributed \mesa{} simulations in $\log_{10} (M_1/M_\odot)$, $\log_{10} (M_2/M_\odot)$, and $\log_{10} (P_\mathrm{orb}/\mathrm{day})$ \edit1{which are kept separate from the regular grid and \ac{AL} data sets until performance evaluation}.

\begin{figure*}
    \centering
    \includegraphics[width=2\columnwidth]{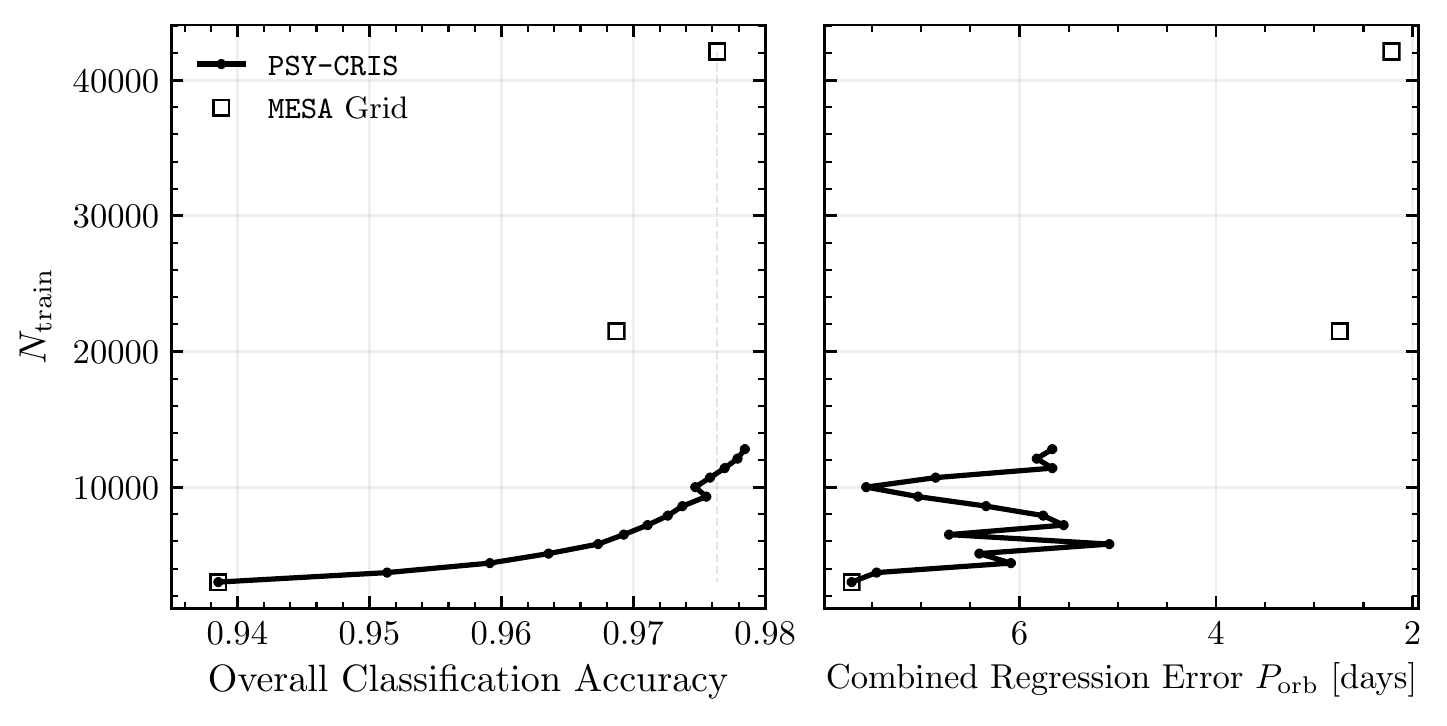}
    \caption{The number of training points needed to achieve a given overall classification accuracy \eqref{eqn:class_accuracy} (left panel) and combined regression error (right panel) in our \ac{CO}--helium-main-sequence \mesa{} test.
    We use a fiducial configuration of \psycris{} with parameters taken from inspecting the synthetic data tests ($\tau=0.5, \beta=2$).
    The data set consists of 5 classes, two of which we consider regression for the final orbital period ($P_\mathrm{orb}$).
    \psycris{} achieves a comparable classification accuracy to the highest density regular grid while using a factor of $\sim4$ fewer \mesa{} simulations.
    In regression, \psycris{} appears to be approaching lower errors comparable to the medium and full density regular grid, but this trend is not strong.
    The per-class performance in classification and regression can be found in \autoref{fig:MESA_per_class_class_perf} and \autoref{fig:MESA_per_cls_regr_perf} respectively. 
    }
    \label{fig:MESA_overall_cls_regr_perf}
\end{figure*}

\begin{figure}
    \centering
    \includegraphics[width=\columnwidth]{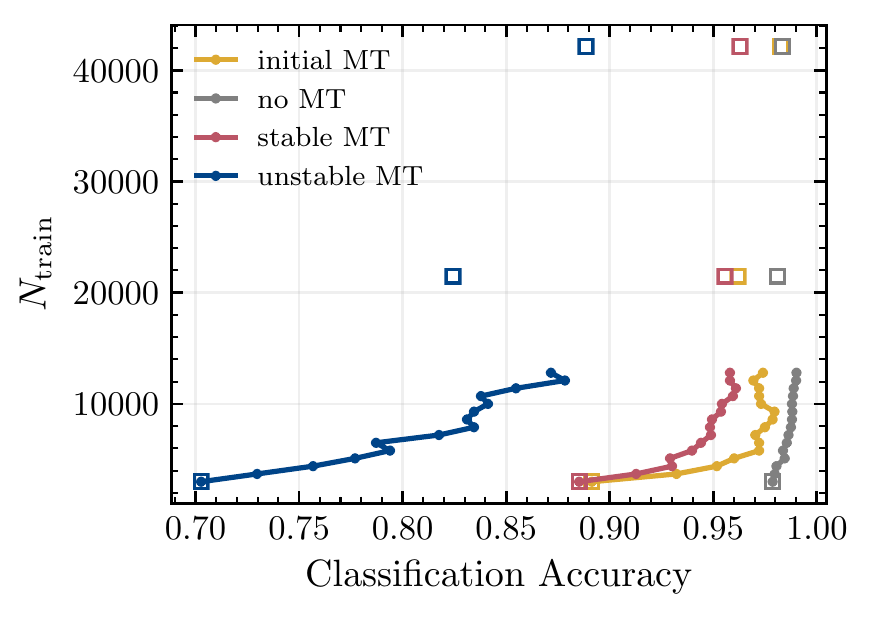}
    \caption{The number of training points necessary to achieve a given per-class classification accuracy (\autoref{eqn:per_class_accuracy}) for the \mesa{} test, using \psycris{} (solid lines) compared to regularly spaced grids (squares).
    We use a fifth class (\texttt{not\_converged}) when running \psycris{} but omit it from performance calculations because it is nonphysical.
    The stable and unstable mass transfer classes have the lowest performance in the data set.
    These classes also fall closer to regions in parameter space where many failed runs occur.
    In all classes we see improvements in classification performance but the most significant gains occur in the unstable mass transfer class.
    }
    \label{fig:MESA_per_class_class_perf}
\end{figure}

\begin{figure*}
    \centering
    \includegraphics[width=2\columnwidth]{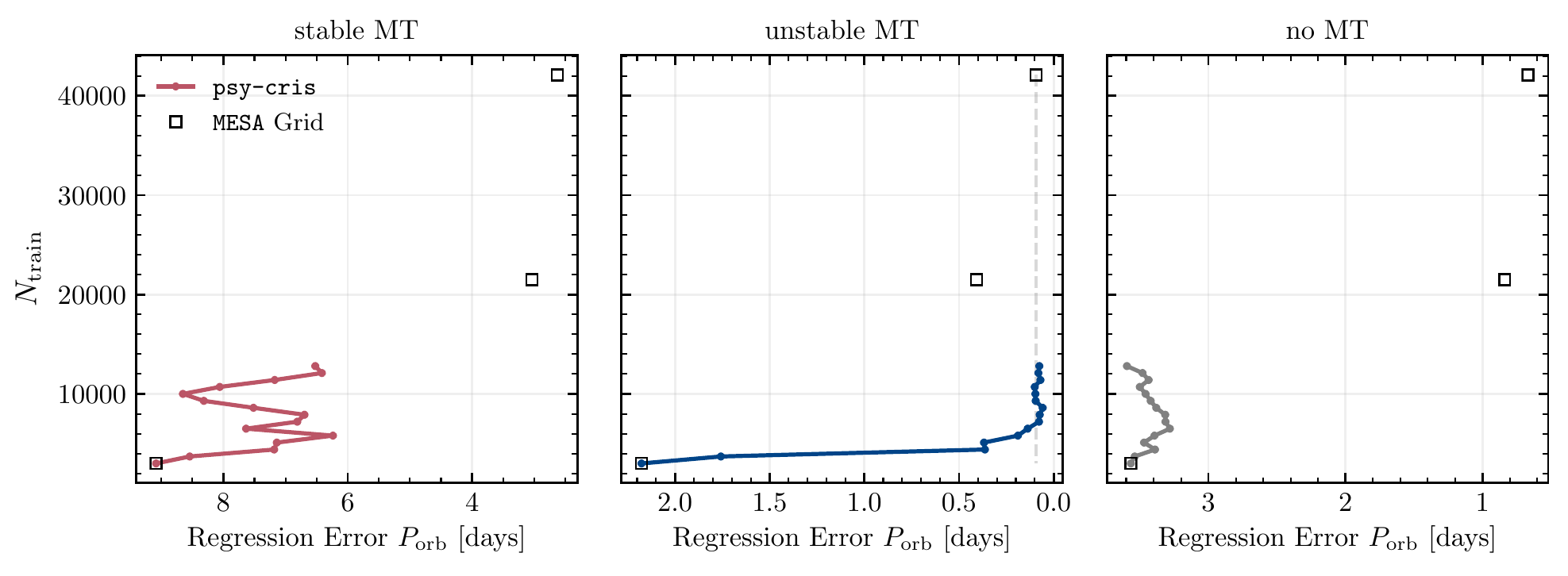}
    \caption{The number of training points necessary to achieve a given regression error per class, defined as the $90$-th percentile of the distribution of absolute differences.
    The squares in each panel show regression performance for regularly spaced \mesa{} simulations at different densities while \psycris{} performance is a solid line.
    Although the simulations were queried by \psycris{} serially, we only show performance calculated every $700$ \mesa{} simulations.
    The \texttt{unstable\_MT} class sees the largest improvements from AL, achieving comparable accuracy to the full density regular grid with a factor of $5.8$ fewer simulations.
    We do not include regression for the \texttt{no\_MT} class during \ac{AL}, explaining the lack of improvement in regression performance.
    The \texttt{stable\_MT}, \texttt{unstable\_MT}, and \texttt{no\_MT} classes make up $35\%$ ($22\%$), $9\%$ ($3\%$), and $30\%$ ($60\%$) \edit1{of the \ac{AL} training data set (random test set), respectively}.
    }
    \label{fig:MESA_per_cls_regr_perf}
\end{figure*}

\begin{figure*}
    \centering
    \includegraphics[width=2\columnwidth]{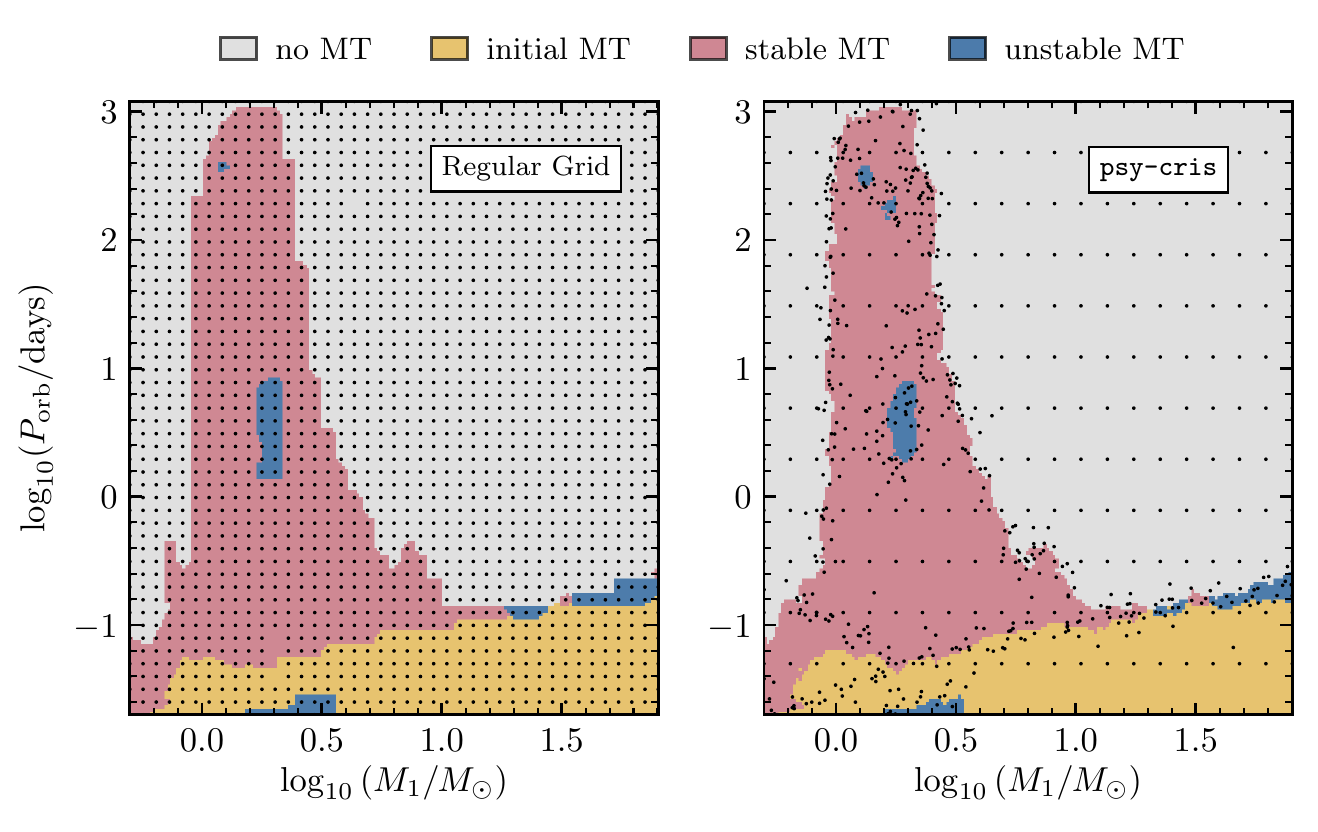}
    \caption{The predicted classification in a slice ($M_2\approx17M_\odot$) of our \mesa{} test using a linear classifier trained on the full density regular grid (left panel) and our \psycris{} \ac{AL} data set (right panel).
    Black dots show the locations of training data within $\pm1/2$ a logarithmic bin width (defined from the full density regular grid) totalling $2{,}004$ points for the regular grid and $750$ points for \psycris{}.
    The starting grid for \psycris{} is a subset of the full density regular grid and contains $3{,}003$ points.
    \psycris{} reproduces the overall structure of the classification space seen in the full density regular grid and resolves finer structure near the class boundaries, while requiring less than half the points.
    There is a small region of \texttt{initial\_MT} simulations labeled as \texttt{unstable\_MT} at $\log(P_\mathrm{orb}/\mathrm{day})\sim-1.6$ and $\log(M_1/M_\odot)\sim0.5$ due to an error in the post-processed output.
    }
    \label{fig:MESA_classification_space}
\end{figure*}

\subsection{\mesa{} test summary}
\label{sec:sub:mesa_test_summary}

In our \mesa{} test we demonstrate how a fiducial, \textit{un-optimized} configuration of \psycris{} performs in the intended application of creating an \ac{AL} grid of \mesa{} simulations for use in \posydon{}.
The process of optimization is highly customized and, hence, not in the scope for this method-demonstration paper.

In \autoref{fig:MESA_overall_cls_regr_perf} we show the overall classification and regression performance between \psycris{} and a regular grid for our \ac{BH}--helium-main-sequence binary simulations.
With \psycris{} we achieve overall classification accuracy comparable to the densest regular grid with $\sim 4$ times fewer \mesa{} simulations.
We also show the combined regression error for the final orbital period $P_\mathrm{orb}$ only, because this is the only quantity we use when considering regression during \ac{AL}.
We see that \psycris{} reaches errors of $\sim 5~\mathrm{day}$ compared to $\sim 2.4~\mathrm{day}$ for the medium-density regular grid.
While we see that the regression accuracy is decreasing, we do not have enough data to identify a strong trend that can be extrapolated to high $N_\mathrm{train}$.
We assess that the variability seen in combined (and per-class) regression error is caused by our \ac{RBF} interpolation, which sets the length scale as the average distance between training data.
We do not to optimize for nor fix the length scale since this is only a demonstration of \psycris{}.
Overall, the performance is consistent with expectations, but to gain a more complete picture, we consider per-class performance for both classification and regression.

In \autoref{fig:MESA_per_class_class_perf} we show the per-class classification accuracy for our \mesa{} test.
The stable- and unstable-mass-transfer classes have the lowest accuracies while the no-mass-transfer class starts at nearly $98\%$ accuracy.
This is due to the relative size of classes in the parameter space, where small classes generally have lower performance.
The unstable-mass-transfer class has the lowest accuracy and sees the largest gains from \ac{AL}, similar to the challenging case of Class 5 in the synthetic data set.
We find that all classes achieve better or comparable per-class classification accuracy with \psycris{} at a lower computational cost than the medium density grid.
The reduction in number of points required for \psycris{} to achieve a comparable per-class accuracy to the medium density regular grid (reduction factor) are given in \autoref{table:MESA_per_class_perf}.
The reduction factor does not translate linearly into CPU-hours saved because not all \mesa{} simulations have the same computational cost.

\begin{deluxetable}{c c c}
    \tablecaption{The factor by which we reduce the required number of training points to achieve the same per-class classification accuracy (\perClassAcc{}) and the same F1-score in the \mesa{} test with \psycris{} compared to the medium-density regular grid in \autoref{fig:MESA_per_class_class_perf}.
    This is simply $N_\mathrm{Grid}/N_\mathrm{AL}$ at a fixed \perClassAcc{} (or F1-score).
    We calculate F1-score as a secondary metric to compare with our \perClassAcc{}, with both metrics showing similar reduction factors.
    All \psycris{} curves except for \texttt{unstable\_MT} for \perClassAcc{}, and \texttt{initial\_MT} for F1-score, achieve comparable accuracy to the highest density regular grid by the end of AL to within $<0.5\%$.}
    \label{table:MESA_per_class_perf}
    \setlength{\tabcolsep}{20pt}
    \tablehead{
    \colhead{} & \multicolumn{2}{c}{Reduction Factor} \\
    \cline{2-3}
    \colhead{Class} & \perClassAcc{} & F1-score
    }
    \startdata
    \texttt{initial\_MT}     & 4.2 & 4.2 \\
    \texttt{no\_MT}          & 4.9 & 3.7 \\
    \texttt{stable\_MT}      & 2.1 & 3.3 \\
    \texttt{unstable\_MT}    & 2.5 & 4.2 \\
    \enddata
\end{deluxetable}

In \autoref{fig:MESA_per_cls_regr_perf} we show the absolute error in final orbital period in days for the stable-, unstable-, and no-mass-transfer classes.
During \ac{AL}, we ignore regression for systems undergoing mass transferring at \acl{ZAMS} (\texttt{initial\_MT}) because we expect these systems to result in stellar mergers early in their evolution, in which case their final orbital period is undefined.
We also ignore regression for systems that do not have Roche-lobe overflow mass transfer (\texttt{no\_MT}) as we assume these systems have comparatively simple evolution, which we are less interested in resolving in this demonstration \psycris{}.

The unstable-mass-transfer class sees the greatest improvements in regression accuracy, achieving comparable performance to the full density regular grid with $5.8$ times fewer \mesa{} simulations.
This large increase in performance is driven by the fact that the unstable-mass-transfer class is least represented in the data (see \autoref{table:MESA_data_sets}), and \ac{AL} necessarily populates underrepresented classes \citep{2007Ertekin}.
Just as in the synthetic data test, we see that each class has different characteristic performances in regression on the final orbital period; the stable-mass-transfer class has errors on the order of days compared to tenths of days for the unstable-mass-transfer class.
The no-mass-transfer class does not achieve any significant improvements in regression, having a nearly constant performance throughout the duration of the \ac{AL} phase.
\edit1{This is consistent with trends from the synthetic data test where disregarding regression during \ac{AL} impacts performance negatively (e.g., $\tau=1$ configurations in \autoref{fig:ad_per_class_regr} showing almost no improvement).}
The stable-mass-transfer class sees some improvements but we do not have enough data to identify strong trends to extrapolate to higher $N_\mathrm{train}$.

Finally, \autoref{fig:MESA_classification_space} shows a slice of our parameter space visualizing the predicted classification after training on the full density regular grid and \psycris{}.
\edit1{The left panel shows the full density regular grid which is missing about five simulations at $\log_{10} (P_\mathrm{orb}/\mathrm{day}) \sim -0.75$ due to errors parsing the \mesa{} simulations.}
The starting grid for \psycris{} is a subset of the full density regular grid and contains $3{,}003$ points.
With \ac{AL} we resolve classification boundaries and even discover a new region of unstable mass transfer systems at $\log_{10} (P_\mathrm{orb}/\mathrm{day}) \sim 2.5$, all while requiring $\lesssim 50\%$ the total number of training points of the regular grid.
This classification performance has been achieved without optimizing \psycris{} for this data set.

\begin{deluxetable}{ l  c c c}
    \tablecaption{The percentage of systems in each class in the random test set, the full density regular grid, and the AL data set (AL points combined with the starting sparse grid).}
    \label{table:MESA_data_sets}
    \setlength{\tabcolsep}{12pt}
    \tablehead{
    \colhead{} & \multicolumn{3}{c}{Data Set Type} \\
    \cline{2-4}
    \colhead{Class} & Random & Grid & AL 
    }
    \startdata
    \texttt{no\_MT}          & 60.0 & 60.3 & 30.4 \\
    \texttt{stable\_MT}      & 20.7 & 20.6 & 35.3 \\
    \texttt{initial\_MT}     & 10.8 & 11.1 & 11.4 \\
    \texttt{not\_converged}  &  5.5 &  4.5 & 14.0 \\
    \texttt{unstable\_MT}    &  3.0 &  3.5 &  8.9 \\
    \enddata
\end{deluxetable}

\section{Discussion}
\label{sec:discussion}

We performed two tests of our \ac{AL} algorithm \psycris{}, on a synthetic data set as well as a real-world application of creating a grid of \mesa{} simulations.

In the synthetic data test, we find that \psycris{} can outperform standard methods of sampling in classification and regression by changing the primary free parameters in our algorithm.
We also find that only focusing on classification or regression during \ac{AL} may significantly hinder performance in the neglected category.
We then use a fiducial configuration of \psycris{} taken from the synthetic data test to construct a dynamic grid of \mesa{} simulations.
In this demonstration, we find \psycris{}, while not optimized for this data set, is able to achieve comparable overall classification accuracy to the full density regular grid with a factor of $\sim 4$ fewer simulations.
We also see significant improvements in per-class regression performance for the unstable-mass-transfer class, achieving comparable performance to the full density regular grid with a factor of $\sim 5.8$ fewer simulations. 
\edit1{However, not all classes see large improvements in performance (e.g., stable-mass-transfer class in regression), so proper optimization of \psycris{} is still necessary and will be explored in future work.}

The \psycris{} algorithm shows promising results for use with \mesa{} grids and is currently being optimized for the \posydon{} project.
For use in other problems, there are several
considerations one should take into account.
We use linear interpolation and \ac{RBF} interpolation from \texttt{scipy} \citep{scipy} to perform classification and regression, respectively.
We also use the same interpolation schemes when evaluating all performance metrics for classification and regression.
There is a free parameter in the \ac{RBF} interpolation, the length scale, which by default is set by the average distance between training data.
We do not attempt to optimize the length scale in the \mesa{} test which is a naive way to perform regression, and we suspect that this causes poor regression performance in general.
In all of our tests we use the same interpolation and classification scheme across all classes, but in practice a more fine-tuned approach is expected to achieve higher accuracies.
A more targeted \ac{AL} regression heuristic may also contribute better performance than shown here.

Concerning AL performance metrics themselves, there are also caveats one should consider.
Although we present common performance metrics in this study, there are multiple ways to characterize performance where each metric may highlight specific features while hiding others.
For example, our per-class classification definition trivially approaches unity for overconfident classifiers.
Imbalanced data sets also present challenges since minority systems may be neglected without affecting cumulative performance metrics significantly \citep{2013Attenberg}.
Therefore, it is important that we explore other performance metrics for \ac{AL} in \posydon{} as we further optimize \psycris{}.

It is impossible to guarantee that our \ac{AL} algorithm, like any \ac{AL} algorithm, will always outperform standard sampling \citep[e.g.,][]{NIPS2004_c61fbef6,2008Settles,2018Yang}.
Our algorithm works for our use case, and it is worth consideration in computationally-challenging problems involving high-dimensional data showing complicated structure and behavior.

\edit1{While \ac{AL} has been shown to increase \ac{ML} performance in high dimensional (${\sim}20$ dimensions, comparable to our parameter space in binary stellar evolution) problems \citep{2019EPJC...79..944C}, further testing is needed to characterize the scaling of \psycris{} in higher than three-dimensional data sets.}
However, we have shown \psycris{} already achieves promising performance in three dimensions, matching the \mesa{} grids being used in \posydon{} \texttt{v1}.
In our demonstration, we are already seeing up to factors of $4$ reduction in the number of simulations to achieve a comparable accuracy with \ac{AL} to high density rectilinear grids.

\edit1{
\subsection{Initial Conditions and Completeness}
The \psycris{} algorithm uses an initial training data set from which it proposes new simulations to be labeled.
However, the starting density of this grid impacts the future evolution of \ac{AL} sampling, as every subsequent query point is based upon the information present in the initial training data set.
For example, if a class does not exist in the initial training set because the starting grid is too sparse, it may never be discovered.
Characterizing all possible outcomes in the parameter space is a problem faced by any discrete sampling method \citep[e.g.,][]{2018ApJS..237....1A,2019MNRAS.490.5228B}, including regular grids and \ac{AL}.
However, in practice, we can use our domain knowledge of binary stellar evolution to check that our initial starting grid covers a reasonable range of physical outcomes, which is what is already done with regular grids at arbitrary densities.
}
\section{Conclusions}
\label{sec:conclusions}
Our aim is to efficiently construct data sets of binary evolution simulations with reliable interpolation accuracy to be used in the population synthesis codes such as \posydon{}.
Regular grids of binary evolution simulations can require prohibitively large computational resources to construct due to the cost per simulation combined with the parameter space we must cover to adequately characterizes binary stellar evolution.
We propose \ac{AL} as a solution for constructing high performance training sets of binary evolution simulations, at a reduced cost compared to standard methods.
We demonstrate our new \ac{AL} algorithm \psycris{} tested on a synthetic data set and a realistic data set of binary evolution simulations with \mesa{}.

In our initial synthetic data set test, we find that \psycris{} can be optimized to outperform standard methods of sampling by varying the free parameters in our model.
We find that focusing just on classification or regression alone may lead to significant losses in performance in the neglected category. 
Therefore it is \emph{essential} to consider both to obtain well-adapted results.
When considering per-class performance, we see major gains from AL in both classification and regression for classes which present challenging classification problems, characterized by extended and narrow shapes in parameter space (Class 5).
We also see low performance occurs for minority classes in imbalanced data sets, as well as a class's presence (or absence) in the initial training set (Class 2).
\psycris{} does indirectly address data imbalances by weighting each class equally (\autoref{table:MESA_data_sets}) but a more active approach may achieve better results.
Different classes will have different characteristic performances based off their size and shape in parameter space, and may also benefit the most from \ac{AL} in challenging cases.

We also demonstrate \psycris{} working with \posydon{} infrastructure to evolve binary systems consisting of a \ac{CO} and helium-main-sequence star with \mesa{}.
We use an unoptimized configuration of \psycris{}, determined from the synthetic data set tests, and find that \psycris{} outperforms the highest density regular grid in overall classification using a quarter the total number of training points.
In regression we see \psycris{} is reducing the regression error but not yet to the level of the medium density regular grid.

Creating the oracle in our \ac{AL} loop was critically important for running simulations with \mesa{}.
Not only does the oracle label simulations automatically, but it ensures that we encounter fewer numerical difficulties that commonly arise in \mesa{}.
In our \mesa{} test, $14\%$ of our \ac{AL} data set consists of failed simulations, which do not terminate normally due to numerical convergence issues (\autoref{table:MESA_data_sets}). 
Minimizing numerical convergence issues in our binary simulations with \mesa{} will only further improve the \ac{AL} results in \posydon{}.

Since each \posydon{} data set will exhibit unique classification and regression challenges, \psycris{} may benefit from further optimization specific to each binary evolution grid.
For instance, one could consider different \ac{AL} metrics such as the classification entropy \citep[e.g.,][]{2017arXiv170302910G,2017Cai,2018Yang}.
Currently, \psycris{} has been designed to consider initial--final quantities alone, but our \mesa{} simulations have time series evolution in between these end points.
Therefore, in the future we are interested in seeing how \psycris{} will respond to the complexity of their time series evolution as a metric for simulation proposals.
Furthermore, we could factor in the varying computing cost of different simulations to pick the most informative points in parameter space per unit CPU time \citep{2010Tomanek,2020Kumar}.

Although we have designed it to optimize our generation of binary evolution grids as part of the \posydon{} code, \psycris{} is a general algorithm that can be applied to data sets with classification and regression outputs.
To be applied directly in another similar problem, one must construct the oracle (which will label and run new simulations) to completely automate the \ac{AL} loop. 
The \psycris{} \ac{AL} approach will work best for optimizing large grids of simulations where the number of simulated points is too large to optimize manually, and brute-force oversampling is too computationally expensive. 
Examples are numerical relativity simulations for building gravitational-wave approximants \citep{Varma:2019csw,2022arXiv220200018H}, or large grids of kilonova models \citep{2022PhRvR...4a3046R}.
The \psycris{} code is open source, and exists as a module in \posydon{} which will be available in the next release of \posydon{}.

\acknowledgements{
We thank Monica Gallegos-Garcia for useful conversation during the development of \psycris{}.
K.A.R.\ is supported by the Gordon and Betty Moore Foundation (PI Kalogera, grant award GBMF8477). 
K.A.R.\ also thanks the LSSTC Data Science Fellowship Program, which is funded by LSSTC, NSF Cybertraining Grant No.\ 1829740, the Brinson Foundation, and the Moore Foundation; their participation in the program has benefited this work.
J.J.A.\ acknowledges support from CIERA and Northwestern University through a Postdoctoral Fellowship.
P.M.\ acknowledges support from the FWO junior postdoctoral fellowship No. 12ZY520N.
C.P.L.B.\ and Z.D.\ acknowledge support from the CIERA Board of Visitors Research Professorship. 
V.K.\ was partially supported through a CIFAR Senior Fellowship and a Guggenheim Fellowship. 
S.B., T.F., K.K., D.M., Z.X. and E.Z. were supported by a Swiss National Science Foundation Professorship grant (PI Fragos, project number PP00P2 176868).
K.K. was partially supported by the Federal Commission for Scholarships for Foreign Students for the Swiss Government Excellence Scholarship (ESKAS No.\ 2021.0277).
Z.X. was supported by the Chinese Scholarship Council (CSC).
The computations were performed at Northwestern University on the Trident computer cluster (funded by the GBMF8477 award). 
This research was supported in part through the computational resources and staff contributions provided for the Quest high performance computing facility at Northwestern University which is jointly supported by the Office of the Provost, the Office for Research, and Northwestern University Information Technology.}

\software{\texttt{NumPy} \citep{numpy}, {\tt SciPy} \citep{scipy}, {\tt matplotlib} \citep{matplotlib}, {\tt pandas} \citep{pandas}, {\tt scikit-learn}, \citep{scikit-learn}, \posydon{} \citep{2022arXiv220205892F}}

\bibliographystyle{aasjournal}
\bibliography{main.bib}

\appendix
\section{Constructing the Synthetic Data Set}
\label{sec:appendix}

In \autoref{sec:synthetic_data_test} we test our \ac{AL} algorithm \psycris{} in a \ac{3D} synthetic data set by comparing classification and regression performance of training sets built with \psycris{} to regularly spaced grids and random sampling.
\edit1{We designed the classification space to contain multiple challenging classification features including a spatially small class, as well as extended and narrow classes throughout the parameter space. 
Our choices were also inspired in part by visual inspection of the binary evolution data sets in \citet{2022arXiv220205892F}, but were otherwise arbitrary to avoid fine tuning.}
We chose a synthetic data set for its cheap computational cost for labeling queries and its deterministic output.
The fiducial range of the synthetic data set is $(-1,1)$ for all dimensions.
The data set contains six unique classes and one regression function, continuous across all classes.

To construct the classification space, we use analytic functions and arbitrary constants (split values) to define binary classification boundaries.
With the first function, two classes are created, and each subsequent function adds one more unique class to the parameter space.
The five functions that define our classification space are given by
\begin{subequations}
 \label{eqn:analytic-classification}
    \begin{align}
    {\psi}_1(x,y,z) &= \displaystyle \left(\frac{x+1}{1.2}\right)^2 + \left(\frac{y-1}{2}\right)^2 + \left(\frac{z}{0.8}\right)^2 - 1 , \\
    {\psi}_2(x,y,z) &= (1-x)^2 + 3(y-x^2)^2 + |z|(x-z)^2 , \\
    {\psi}_3(x,y,z) &= \sin( x-0.25 ) + y z  , \\
    {\psi}_4(x,y,z) &= |x^3 - y^3|\exp[-z] , \\ 
    {\psi}_5(x,y,z) &= \left(\frac{x-0.5}{0.6}\right)^2 + \left(\frac{y-0.5}{0.2}\right)^2 + \left(\frac{z-0.85}{0.3}\right)^2 - 0.2^2, \label{eqn:sub:y5}
    \end{align}
\end{subequations}
where $x$, $y$, and $z$ are inputs in Cartesian coordinates.
We combine each function and their associated split values into inequalities such that any input will be mapped to True or False for each function.
We provide a pseudo-code function for the classification of an arbitrary point in our domain,
\begin{verbatim}
    def classification(x,y,z):
        if psi_5(x,y,z) < 0.3:
            return "Class 2"
        elif psi_4(x,y,z) > 1.8 and z < 0:
            return "Class 4"
        elif psi_1(x,y,z) > 1 and psi_2(x,y,z) > 2.5:
            return "Class 1"
        elif psi_2(x,y,z) > 2.5:
            return "Class 6"
        elif psi_3(x,y,z) > -0.5:
            return "Class 3"
        else:
            return "Class 5"
\end{verbatim}
where methods \texttt{psi\_1} through \texttt{psi\_5} correspond to the functions in \autoref{eqn:analytic-classification}.
Finally, we have the regression function $\psi_6$ which is a \ac{3D} sinusoid, spherically symmetric about the origin.
We chose to use the same function for all classes even though \psycris{} treats each class' regression output separately.
\begin{equation}
    \psi_6(x,y,z) = \sin\left( 8\pi \sqrt{x^2+y^2+z^2}/3 
    \right) .
    \label{eqn:analytic-regression}
\end{equation}

\begin{figure*}
    \centering
    \includegraphics[width=\columnwidth]{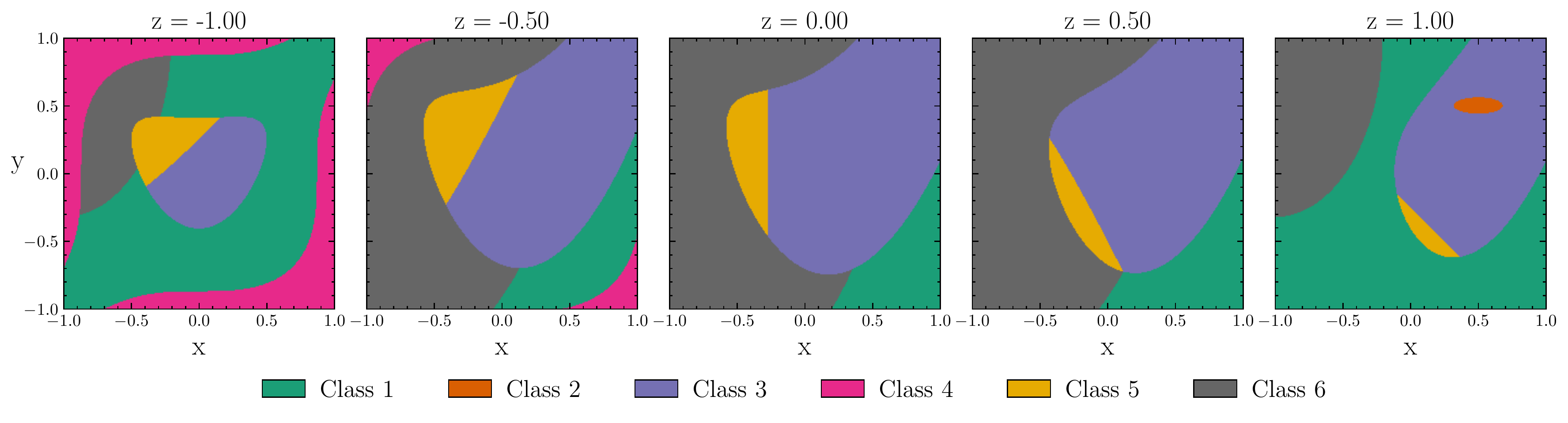}
    \caption{The true classification space for the synthetic data set, constructed with the analytic functions provided in \autoref{eqn:analytic-classification}.
    The regression output from \autoref{eqn:analytic-regression} is defined in the same range, and together they make up the synthetic data set.
    }
    \label{fig:synth_data_3D}
\end{figure*}

\end{document}